\documentclass[preprint2]{proto}
\usepackage{times}
\usepackage{natbib}

\begin{document}

\title{\textbf{\LARGE Volatiles in protoplanetary disks}}

\author{\textbf{\large Klaus M. Pontoppidan}}
\affil{\small\em Space Telescope Science Institute}
\author{\textbf{\large Colette Salyk}}
\affil{\small\em National Optical Astronomy Observatory}
\author{\textbf{\large Edwin A. Bergin}}
\affil{\small\em University of Michigan}
\author{\textbf{\large Sean Brittain}}
\affil{\small\em Clemson University}
\author{\textbf{\large Bernard Marty}}
\affil{\small\em Universit{\'e} de Lorraine}
\author{\textbf{\large Olivier Mousis}}
\affil{\small\em Universit{\'e} de Franche-Comt{\'e}}
\author{\textbf{\large Karin I. {\"O}berg}}
\affil{\small\em Harvard University}

\begin{abstract}
\baselineskip = 11pt
\leftskip = 0.65in 
\rightskip = 0.65in
\parindent=1pc
{\small 
\noindent 

Volatiles are compounds with low sublimation temperatures, and they make up most of the condensible mass in typical planet-forming environments. They consist of relatively small, often hydrogenated, molecules based on the abundant elements carbon, nitrogen and oxygen. Volatiles are central to the process of planet formation, forming the backbone of a rich chemistry that sets the initial conditions for the formation of planetary atmospheres, and act as a solid mass reservoir catalyzing the formation of planets and planetesimals. Since Protostars and Planets V, our understanding of the evolution of volatiles in protoplanetary environments has grown tremendously. This growth has been driven by rapid advances in observations and models of protoplanetary disks, and by a deepening understanding of the cosmochemistry of the solar system. Indeed, it is only in the past few years that representative samples of molecules have been discovered in great abundance throughout protoplanetary disks (CO, H$_2$O, HCN, C$_2$H$_2$, CO$_2$, HCO$^+$) -- enough to begin building a complete budget for the most abundant elements after hydrogen and helium. The spatial distributions of key volatiles are being mapped, snow lines are directly seen and quantified, and distinct chemical regions within protoplanetary disks are being identified, characterized and modeled. Theoretical processes invoked to explain the solar system record are now being observationally constrained in protoplanetary disks, including transport of icy bodies and concentration of bulk condensibles. The balance between chemical {\it reset} -- processing of inner disk material strong enough to destroy its memory of past chemistry, and {\it inheritance} -- the chemically gentle accretion of pristine material from the interstellar medium in the outer disk, ultimately determines the final composition of pre-planetary matter. This chapter focuses on making the first steps toward understanding whether the planet formation processes that led to our solar system are universal. 
 \\~\\~\\~}

\end{abstract}  

\section{\textbf{Introduction}}
The study of the role of ices and volatile compounds in the formation and evolution of planetary systems has a long and venerable history. Up until the late 20th century, due to the lack of knowledge of exo-planetary systems, the solar system was the only case study. Consequently, some of the earliest pieces of firm evidence for abundant ices in planetary systems came from cometary spectroscopy, showing the photodissociation products of what could only be common ices, such as water and ammonia \citep{Whipple50}. Following early suggestions by e.g., \cite{Kuiper53}, the presence of water ice in the rings of Saturn \citep{McCord71} and in the Jovian satellite system \citep{Pilcher72} was confirmed using infrared spectroscopy. The characterization of ices in the outer solar system continues to this day \citep{Mumma11,Brown12}. Similarly, the presence of ices in the dense interstellar medium -- the material out of which all planetary systems form -- has been recognized since the 1970s and conjectured even earlier. With the advent of powerful infrared satellite observatories, such as the Infrared Space Observatorory (ISO) and the Spitzer Space Telescope, we know that interstellar ices are commonplace and carry a large fraction of the solid mass in protostellar environments \citep{Gibb04,Oberg11}.

The next logical investigative steps include the tracking of the volatiles as they take part in the formation and evolution of protoplanetary disks --- the intermediate evolutionary stage between the interstellar medium and evolved planetary systems during which the planets actually form. This subject, however, has seen little progress until recently. A decade ago, volatiles in protoplanetary disks were essentially beyond our observational capabilities and the paths the volatiles take from the initial conditions of a protoplanetary disk until they are incorporated into planets and planetesimals were not well understood. What happened over the past decade, and in particular since the conclusion of PPV, is that our observational knowledge of volatiles in protoplanetary disks has been greatly expanded, opening up a new era of comparative cosmochemistry. That is, the solar nebula is no longer an isolated case study, but a data point -- albeit an important one -- among hundreds. The emerging complementary study of volatiles in protoplanetary disks thus feeds on comparisons between the properties of current-day solar system material and planet-forming gas and dust during the critical first few million years of the development of exo-planetary systems. This is an area of intense contemporary study, and is the subject of this chapter.

\subsection{Emerging questions}

Perhaps one of the most central questions in astronomy today is whether our solar system, or any of its characteristics, is common or an oddity? We already know that most planetary systems have orbital architectures that do not resemble the solar system, but is the chemistry of the solar system also uncommon? Another matter is the degree to which volatiles are {\it inherited} from the parent molecular cloud, or whether their chemistry is {\it reset} as part of typical disk evolution. That is, can we recover evidence for an interstellar origin in protoplanetary and planetary material, or is that early history lost in the proverbial furnace of planet formation? There is currently an apparent disconnect between the cosmochemical idea of an ideal condensation sequence from a fully vaporized and hot early phase in the solar nebula, supported by a wealth of data from meteoritic material, and the astrophysical idea of a relatively quiescent path of minimally processed material during the formation of protoplanetary disks. The answer is likely a compromise, in which parts of the disk are violently reset, while others are inherited and preserved over the lifetime of the central star, and where most regions show some evidence for both, due to a variety of mixing processes. In this chapter, we will consider the solar nebula as a protoplanetary disk among many others -- and indeed call it the solar protoplanetary disk. We will discuss the issue of inheritance versus reset, and suggest a division of disks into regions characterized by these very different chemical circumstances. 

\subsection{Defining protoplanetary disks and volatiles}
The term {\it protoplanetary disk} generally refers to the rotationally supported, gas-rich accretion disk surrounding a young pre-main sequence star. The gas-rich disk persists during planetesimal and giant planet formation, but not necessarily during the final assembly of terrestrial planets. During the lifetime of a protoplanetary disk, both solid- and gas-phase chemistry is active, shaping the initial composition of planets, asteroids, comets and Kuiper belt objects. Many of the chemical properties and architecture of the current day solar system were set during the nebular/protoplanetary disk phase, and are indeed preserved to this day, although somewhat obscured by dynamical mixing processes during the later debris disk phase. 

Most of the mass in protoplanetary disks is in the form of molecular hydrogen and helium gas. Some of the solid mass is carried as dust grains, mostly composed of silicates and with some contribution of carbon-dominated material. This material is generally {\it refractory}, that is, very high temperatures ($\gtrsim$ 1000\,K) are needed to sublimate it. Only in a very limited region, in a limited period of time or under unusual circumstances are refractory grains returned to the gas phase. The opposite of refractory is {\it volatile}. In the context of this chapter, disk volatiles are molecular or atomic species with relatively low sublimation temperatures ($<$ a few 100\,K) that are found in the gas phase throughout a sigificant portion of a typical disk under typical, quiescent disk conditions\footnote{In cosmochemistry the use of the words refractory and volatile is somewhat different. While they are still related to condensation temperature, their use is usually reserved for atomic elements, rather than molecules, for interpreting elemental abundances in meteorites.}. It is also possible to define a sub-class of volatile material that includes all {\it condensible} species -- that is, the class of volatiles that are found in both their solid and gaseous forms in significant parts of the disk, but which excludes species that never condense in bulk, such as H$_2$. It is the ability of condensible volatiles to relatively easily undergo dramatic phase changes that lead to them having a special role in the evolution of protoplanetary disks and the formation of planets. An example of a volatile that is not considered condensible is the dominant mass compound, H$_2$, while one of the most important condensible species is water.

\section{\textbf{The solar nebula as a volatile-rich protoplanetary disk}}
\label{SolarNebula}
\vspace{0.15cm}

The solar system was formed from a protoplanetary disk during a time frame of 2-3 million years, spanning ages of at least 4567-4564 Myr -- a number known to a high degree of precision, thanks to accurate radiometric dating of primitive meteoritic material formed by gas-condensation processes \citep{Scott07}. The composition of solar system bodies, including primitive material in meteorites, comets, and planets provide a detailed window into their formation. The primordial elemental composition of the solar protoplanetary disk has been systematically inferred from measurements of the solar atmosphere, generally under assumptions of efficient convective mixing. Any compositional or isotopic differences measured in primitive solar system materials, such as comets and meteorites, can therefore be ascribed to processes taking place during the formation and evolution of the solar system, generally after the formation of the Sun.

One of the initial models to account for the volatility patterns of meteorites is one where the solar disk started hot ($>$ 1400 K), such that all material was vaporized.  As the gas cooled, the elements formed minerals governed by the condensation sequence in thermodynamic equilibrium \citep[e.g.][]{grossman72, wh93, eb00}, with some level of incompletion due to grain growth and disk gas dispersal \citep{wc74, cassen96}. Condensation models predict that Calcium-Aluminum compounds would be the first elements to condense \citep{lord65, grossman72}, a proposition that is consistent with Calcium-Aluminum Inclusions (CAI) being the oldest material found in the solar system \citep{amelin02}. 

The widespread presence of presolar grains, albeit in extremely dilute quantities \citep{zinner98}, the high levels of deuterium enrichments seen in Oort cloud comets and meteorites \citep{Mumma11, alexander12}, and the presence of highly volatile CO in comets, presents the contrary perspective that at least part of the nebula remained cold or represents material provided to the disk at later cooler stages. It is with these fundamental contradictions in mind that we strive to understand the solar system and general protoplanetary disks under a common theoretical umbrella. 

\subsection{The bulk abundances of carbon, oxygen and nitrogen}
\begin{figure}[ht!]
\begin{center}
\includegraphics[width=8cm]{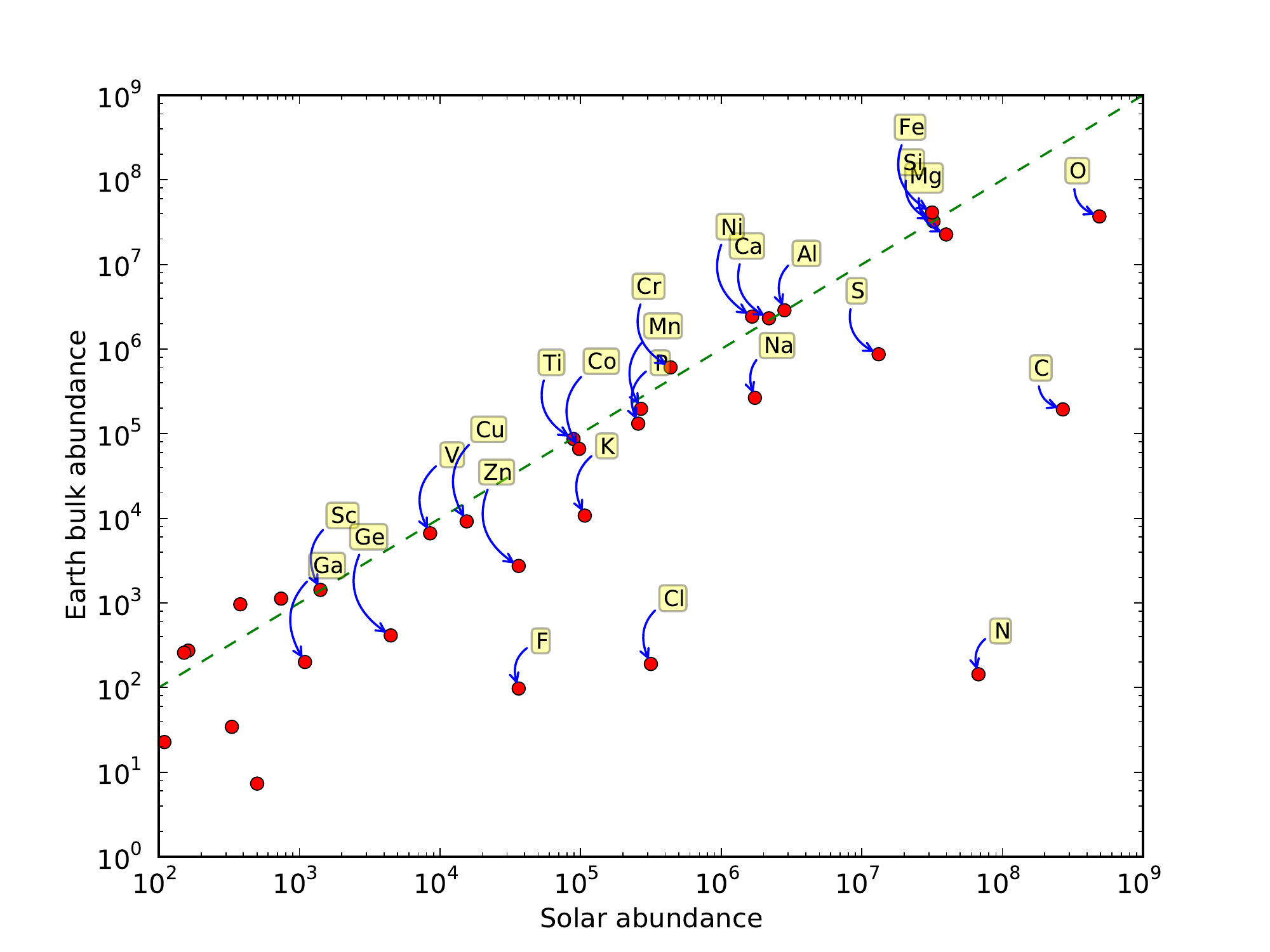}
\caption{Solid Earth bulk elemental abundances \citep{Allegre01} compared to solar abundances \citep{Grevesse10}. The solar abundances are normalized to an H abundance of $10^{12}$. For the Earth, the silicon abundance is scaled to that of the Sun. All abundances are in units of relative number density of nuclei. The dashed line guides the eye to the 1:1 ratio. For instances, the solar abundances of carbon, oxygen and nitrogen are $10^{8.43}$, $10^{8.69}$ and $10^{7.83}$, respectively.}
\label{fig:earth_abundance}      
\end{center}
\end{figure}

\begin{figure}[ht!]
\begin{center}
\includegraphics[width=7.5cm]{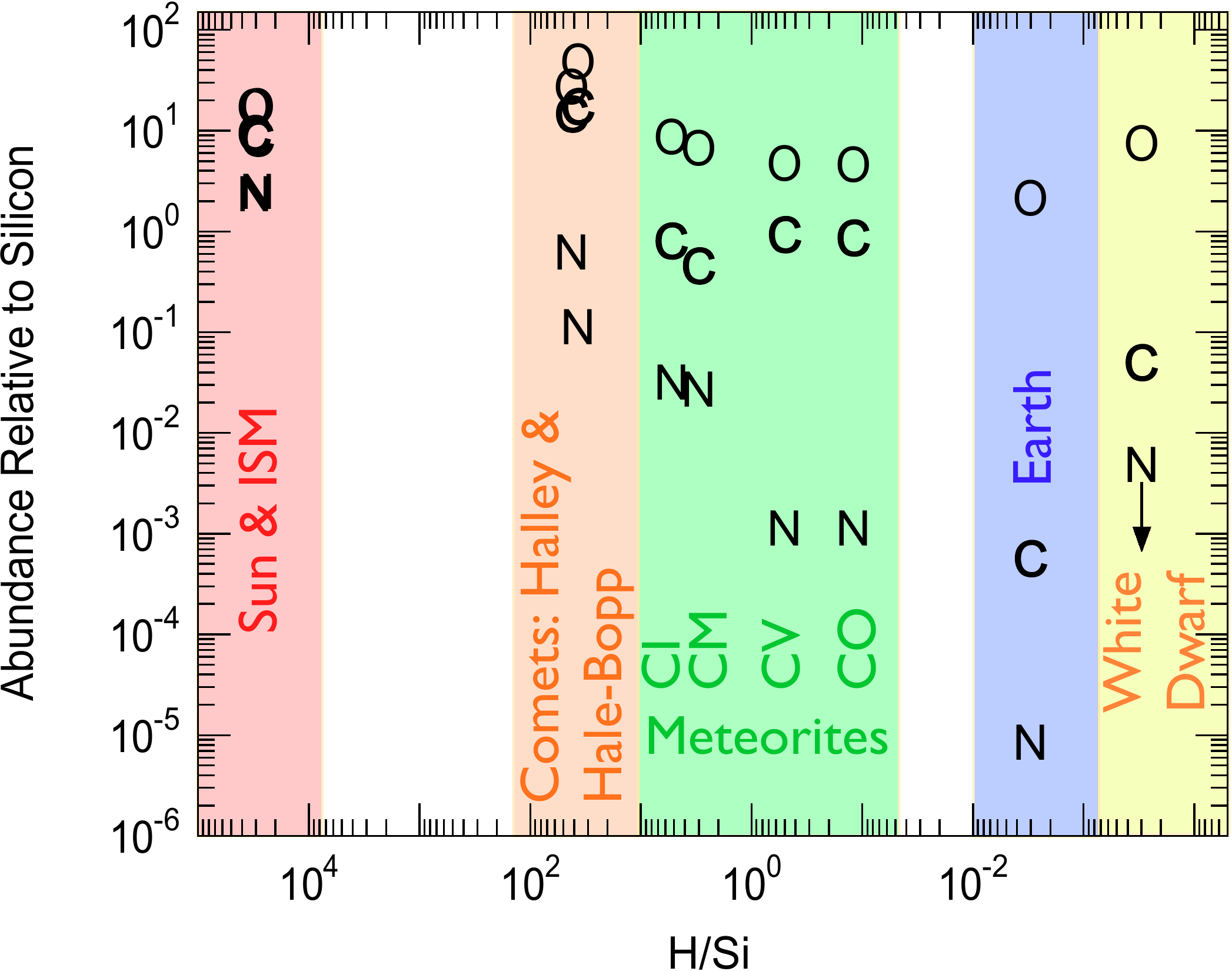}
\caption{The relative CNO abundances in the solar system. The abundances are computed relative to Si, the primary refractory element in the solar system and interstellar space. The abundances are shown as a function of the H/Si ratio which separates the various bodies. The figure is taken and updated from \citet{lbn10,Geiss87}.  The white dwarf values are for GD 40 \citep{Jura12} and are provided with an arbitrary H/Si ratio (since this is unknown).}
\label{fig:CONreltoSi}      
\end{center}
\end{figure}

For instance, consider the bulk abundances of common elements in the Earth, and in particular of carbon, oxygen and nitrogen, which are the main building blocks, along with hydrogen, of volatile chemistry. With a surface dominated by carbon-based life, coated by water, with a nitrogen-rich atmosphere, our planet is actually a carbon-, nitrogen- and water-poor world, compared to solar abundances and, presumably, to the abundance ratios of the molecular cloud out of which the solar system formed. This observable is illustrated in Figure\ \ref{fig:earth_abundance}, where the Earth's relative elemental abundances are compared to those of the Sun. While the refractory elements are essentially of solar abundances, others are depleted by orders of magnitude, most prominently carbon and nitrogen, and to a lesser degree, oxygen. Going beyond the Earth, C, N and O abundances in various solar system bodies are compared (Earth, meteorites, comets, and the Sun) in Figure \ref{fig:CONreltoSi}. 

A great amount of information about the relative distribution of volatiles in the solar nebula is contained in these plots. Specifically, they indicate the volatility of the chemical compounds carrying the bulk of each element: If an element is highly depleted, most of that element must either have been in a form too volatile to condense at the location and time that a specific solar system body formed, or the material was subjected to subsequent, possibly transient, heating at a later time while still in a form where outgassing and loss could be efficient. In other words, while the plot will not directly tell us what the main elemental carriers were, it will provide us with some of their basic physical-chemical properties. In the case of the Earth, we can see that many elements, and not only C, N and O, are depleted to some extent, giving rise to the idea that the Earth formed from material exposed to temperatures well above 150\,K, and going as high as 1000\,K \citep{Allegre01}.

The ISM and the Sun represent the total amount of material available. Beyond the Sun, most bodies in the solar system are hydrogen-poor. This reflects the fact that molecular hydrogen is the most volatile molecular compound present in a protoplanetary disk and is never efficiently incorporated into solids anywhere in the solar system. Comets formed in the coldest regions of the nebula and incorporate the largest amount of hydrogen, mostly trapped in the form of water. In contrast, the rocky planetesimals that formed close to the Sun contain successively less water, and therefore less hydrogen.

{\bf The oxygen} content in the solar system is also, to a large degree, a story of water. Oxygen is generally the least depleted element among the CNO trifecta across the solar system, indicating that a relatively more refractory compound carried much of the oxygen. A likely candidate for this carrier is silicates (mostly oxides built with the anions SiO$_4^{4-}$ or SiO$_3^{2-}$). Indeed, comets have the highest O/Si ratio, above that even of the Sun, although there is some uncertainty in the calculation of silicon abundances in comets as these depend on an assumed mass opacity \citep{min05}.  However, one value comes from {\em in situ} measurements of Comet Halley by the Soviet Vega-1 spacecraft \citep{jessberger88}, and supports this general picture. Thus, comets incorporated oxygen in the form of silicates, as well as CNO ices. Rocky bodies also contain significant amounts of oxygen, but below the total atomic nebular reservoir, by factors of 2-10. This allows a rough estimate of the fraction of oxygen in volatiles versus refractory compounds -- a few volatile oxygen atoms for each refractory oxygen atom. Furthermore, it demonstrates that in the terrestrial planet forming regions of the nebula (inside the snow line), most oxygen was in a gaseous form.  Overall, the variation in bulk oxygen abundance throughout the solar system suggests that the main oxygen carrier was a {\it condensible} species --- frozen out in the outer disk, yet in the gas-phase in the inner disk.  From these considerations, likely candidates are H$_2$O, CO$_2$ and CO. 

{\bf The bulk carbon} abundance shows a stronger signal than that of oxygen. As we traverse the solar system, from comets over the asteroids to the Earth, we discover a gradient in the relative amounts of condensed carbon, culminating in a carbon depletion of three orders of magnitude in the Earth. The caveat is that the Earth's C/Si ratio is estimated from the mantle \citep{Allegre01}, but the Earth's total carbon content is highly uncertain as there might be deep reservoirs of carbon in the lower mantle and/or core \citep{wood93, dh10}.   However, meteorites have been posited as tracers of the Earth's starting materials and even unprocessed carbonaceous chondrites have nearly an order of magnitude carbon deficiency relative to the total amount that was available in the nebula. The likely inference is that within the inner nebula volatile carbon must have been predominantly in a volatile, gaseous form, in contrast with the ISM, where much, perhaps most, of the carbon is sequestered as graphite or other highly refractory forms. This so-called {\it carbon deficit problem} will be discussed further in \S\ref{carbon_def}. The carbon deficit in planetesimals is contrasted with a carbon enhancement in the giant planet atmospheres; the C/H in the Jovian atmosphere is enhanced by a factor 3.5 over solar \citep{Niemann96,Wong04}, while the Saturnian ratio is 7 times solar \citep{Flasar05}. This is consistent with the giant planets forming outside the condensation front of a primary volatile carbon carrier.  

The carbon that made it to the Earth was likely delivered after the protoplanetary disk phase, but during the bulk Earth formation, and there is significant evidence that it is mostly of chondritic origin \citep{Marty12, Albarede13}. Carbonaceous chondrites are composed primarily of silicates, but contain up to 10-20\% equivalent water in the form of hydrated silicates, and a few percent of carbon, by mass \citep{Scott07}. While strongly depleted in bulk carbon relative to solar abundances, the remaining chondritic carbon (and nitrogen) is mostly carried by refractory organics. The origin of chondritic organics is often traced by the isotopic ratios of noble gases (Ar, Ne, Xe, ...), trapped in an ill-defined organic phase (called ``Q''; \citealp{Ott81,Marrocchi05}). Present in very small fractions ($<1$\% by mass), chondritic noble gases generally have a recurring isotopic signature, strongly fractionated relative to that of the Sun and the solar protoplanetary disk (chondritic noble gases are heavy). The isotopic signature in noble gases in the Earth's atmosphere is consistent with a mixture that is 90\% chondritic and at most 10\% solar \citep{Marty12}. The chondritic noble gas fractionation has been reproduced in the laboratory during gas-solid exchanges under ionizing conditions \citep{Frick79}, consistent with entrapment during photochemical formation of the organics from ices in the solar disk \citep{Jenniskens93,Gerakines96,Oberg09}. Another way of correctly fractionating noble gases is by selective trapping in forming ices. \cite{Notesco99} found experimentally that noble gases are enriched according to their isotopic mass ratios ($\sqrt{m_1/m_2}$) during the formation of amorphous water ice. That is, a similar effect can be generated as a purely thermal effect. In summary, there is significant evidence that the Earth's carbon is delivered with a volatile reservoir. That said, it is not clear whether the delivery vector was from a chondritic reservoir, as opposed to a cometary reservoir. There is support for chondritic delivery, but an unambiguous identification was for a long time not possible due to a lack of accurate measurements of noble gases in comets. Recently, samples returned by the stardust mission from comet 81P/Wild 2 showed similar noble gas signatures as those of chondrites, indicating that the noble gas hosts may be of similar origin in comets and chondrites \citep{Marty08}.

{\bf Nitrogen} is the most depleted element in our selection. In the Earth, it is depleted by 5 orders of magnitude. While the Earth's atmosphere is dominated by N$_2$, this represents about half of the Earth's nitrogen \citep{Marty95}. In comparison, the total mass of the atmosphere is a bit less than 1 ppm of the total mass of the Earth. While the absolute abundance of nitrogen is uncertain in comets, it is clear that it is highly depleted relative to the Solar value. The best estimates come from the Halley {\em in situ} measurements \citep{wte91} as this accounts for N incorporated in the coma as evaporated ices and in the refractory component. The Hale-Bopp estimate \citep[taken from the summary of][]{Mumma11}, which has the lower N/Si ratio in Figure \ref{fig:CONreltoSi}, only includes the ices and is therefore a lower limit. One interesting aspect is that \citet{wte91} estimates that Comet Halley carried most of its nitrogen (90\%) in a carbonaceous (CHON) refractory component. As in the other heavy element pools, there is a gradient in the N/Si ratio as we approach the terrestrial planet forming region with both meteorites and the Earth exhibiting large nitrogen depletions, relative to the total amount available. Clearly, the vast majority of the nitrogen resided in the gas in the solar nebula in a highly volatile form, which could be NH$_3$, N$_2$ and/or N. In this context, laboratory measurements suggest that both CO and N$_2$ have similar binding strengths to the grain surface \citep{oberg_n2}. Thus the difference between the presence of CO in comets and the relative lack of N$_2$ stands out. Note, however, that in the Jovian atmosphere, nitrogen in the form of NH$_3$ is enhanced relative to solar by a factor 3 \citep{Wong04}, similar to that of carbon, and similarly indicating that the Jovian core formed outside of the primary nitrogen condensation front.

\subsection{Origin of the carbon deficit}
\label{carbon_def}
As we have seen in Figure \ref{fig:CONreltoSi}, solids in the terrestrial region of the solar system, as represented by the Earth and meteorites, have far less elemental carbon than what was available, on average, in the solar protoplanetary disk. The implication is that the carbon was either entirely missing in the terrestrial region, or it was in a volatile form, not available for condensation. In stark contrast, the dense interstellar medium sequesters $\sim 60$\% of the carbon in some unknown combination of amorphous carbon grains,  large organics, or, perhaps, as polycyclic aromatic hydrocarbons (PAHs) \citep{ss96, draine03}. If terrestrial worlds formed directly from solid grains that are supplied to the young disk via collapse then they would have 2 times more carbon than silicon by mass. This is strikingly different from the observed ratio of 1-2\% of C/Si by mass \citep{Allegre01} and essentially all of the primordial, highly refractory, carbon carriers must be destroyed prior to terrestrial planet formation to account for the composition of the Earth {\em and} carbonaceous chondrites. 

A similar signature is known to exist in exoplanetary systems from measurements of asteroidal material accreted onto white dwarfs \citep{jura06}.  A small fraction of white dwarfs show lines from a wide range of heavy elements. Since any heavy elements are rapidly depleted from white dwarf photospheres by settling, their presence indicates that they must have been recently accreted \citep{paq86}. In such {\it polluted} white dwarfs, the C/Fe and C/Si abundance ratios are well below solar and consistent with accretion of asteroidal material deficient in carbon \citep{jura08, gansicke12, Farihi13}. This suggests that the Earth's missing carbon and the carbon deficit in the inner solar system are a natural outcome of star and planet formation, that is, the process leading to a strong carbon deficit is a universal property of protoplanetary disks.

Theoretical models suggest the carbon grains can be destroyed via oxidation either by OH at the inner edge of the accretion disk \citep{gail02} or by oxygen atoms created by photodissociation of oxygen-carrying volatiles, such as water, on disk surfaces exposed to ultraviolet radiation from the central star (or other nearby stars) \citep{lbn10}. Although the details vary depending upon the uncertain carrier of carbon grain absorption in the interstellar medium, the destruction of the grains could produce CO or CO$_2$ (graphite/amorphous carbon) or, to a lesser degree, hydrocarbons with oxygen from PAHs \citep{gail02, lbn10, Kress10}. The theoretical implication is that CO or CO$_2$ are the likely dominant carriers of carbon in the inner disk, rather than either solid-phase carbon grains or gas-phase molecules like CH$_4$, and that the sum of these species should be close to the solar carbon abundance. 

\subsection{The solar system snow line}
We know that the distribution of planets in the solar system is skewed, with low-mass terrestrial planets residing inside of 3 AU and massive gas and ice giant planets orbiting beyond this radius, and it has long been thought that this is related to the presence of a water ice condensation front (the so-called ``snow line'') being located at a few AU \citep{Hayashi81}. However, the present-day distribution of water in the asteroid belt is somewhat ambiguous, and the location of the end-stage snow line in the solar disk may be preserved only as a fossil. The lifetime of water ice on the surfaces of all but the largest asteroids is short-lived inside of 5 AU \citep{Lebofsky80,Levison97}. Exceptions are 1 Ceres, which has some amount of surface water ice \citep{Lebofsky81}, and 24 Themis \citep{Campins10,Rivkin10}. In both cases, this may be evidence for subsurface ice revealed by recent activity. For most asteroids, the water content has been modified by heating within a few Myr of disk formation. This heating led to the melting of interior ice, followed by aqueous alteration and the formation of hydrated silicates that are stable up to the current age of the solar system.

The direct methods searching for asteroid surface water in the form of hydrated silicates have encountered some interpretational difficulties. Hydrated silicates on the surfaces of outer belt asteroids with semi-major axes in the range 2.5- 5.2\,AU do not show clear evidence for a snow line \citep{Jones90}, and their presence may therefore be more a tracer of local thermal evolution than the primordial water ice distribution \citep{Rivkin02}. That is, they may be a tracer of planetesimals that were exposed to transient heating strong enough to melt water and cause the formation of the hydrated silicates, rather than the presence of water per se. 

The localization of the solar system snow line relies on the measured water content of meteorites, coupled with spectroscopic links to taxonomic asteroid classes \citep{Chapman75,Chapman96}. Carbonaceous chondrites (CI and CM) contain up to 5-10\% of water, by mass \citep{Kerridge85}, and have been associated with C-type asteroids, which have semi-major axes predominantly beyond 2.5\,AU \citep{Bus02}. Conversely, ordinary chondrites are relatively dry (0.1\% of water, by mass) \citep{McNaughton81} and are associated with S-type asteroids, which are found within 2.5\,AU \citep{Bus02}. The different asteroid types, however, are radially mixed, and any present-day water distribution is not as sharp as the snow line once was. This is likely, at least in part, related to dynamical mixing of planetesimals as described by the Nice and Grand Tack models \citep{Walsh11}.

In summary, the study of the solar nebula snow line is associated with significant uncertainties due to a lack of in situ measurements of asteroids and smearing out by dynamical mixing that astronomical measurements of snow lines in protoplanetary disks may provide important new constraints on this critical feature of a forming planetary system.

\subsection{Isotope fractionation in solar system volatiles}
\label{sec:fractionation}
\begin{figure}[ht!]
\begin{center}
\includegraphics[angle=0,scale=1.25]{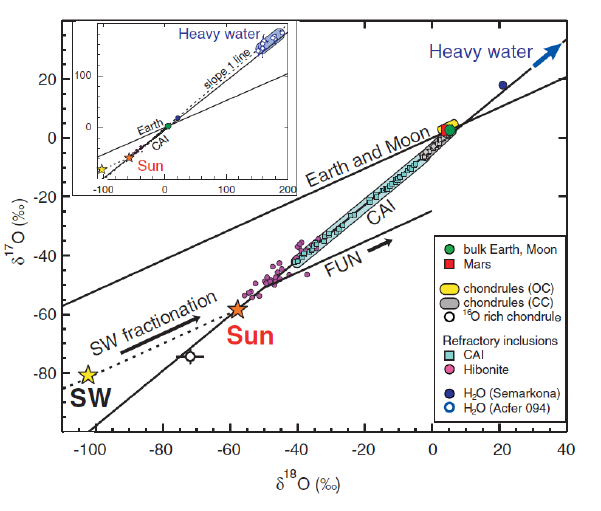}
\caption{The three-isotope plot of oxygen in the solar system, showing the fractionation line of meteorites along a mass-independent slope-1 line. The solar point is a recent addition from Genesis samples of the solar wind.  This point, labeled ``Sun'', is deduced from the sample measurements, labeled ``SW'', by assuming a correction for mass-dependent fractionation in the solar wind. The inset figure shows an extended view highlighting the very $^{16}$O depleted samples of heavy water in the carbonaceous chondrite Acfer 094.  Figure originally from \citet{McKeegan11}.}
\label{fig_O}      
\end{center}
\end{figure}

One of the classical methods for investigating the origin of solar system materials is the measurement and comparison of accurate isotopic ratios. This is in particular true for CNOH as the main volatile building blocks, and strong fractionation patterns are indeed observed in the bulk carriers of these elements in the solar system. The most important systems include the deuterium/hydrogen (D/H) fraction, the relative amounts of the three stable isotopes of oxygen, $^{16}$O, $^{17}$O and $^{18}$O, as well as the ratio of the two stable nitrogen isotopes, $^{14}$N/$^{15}$N. Since these are abundant species -- in particular relative to the many rare refractory systems traditionally used in cosmochemistry, they are also accessible, albeit with difficulty, to astronomical observations of the interstellar medium and protoplanetary disks. These elements show great isotopic heterogeneity in various solar system reservoirs, in sharp contrast to more refractory elements, of which most are isotopically homogenized at the 0.1\% level --- including elements with somewhat lower condensation temperatures, such as Cu, K and Zn \citep{Zhu01, Luck01}. Of particular interest is meteoritic oxygen, which scatters along the famous mass-independent fractionation line in a three-isotope plot (\citealp{Clayton73}; see Figure\ \ref{fig_O}). The $^{14}$N/$^{15}$N ratio exhibits large variations of up to a factor 5 among different solar system bodies \citep{Briani09}. The ubiquitous strong fractionation of the CNOH systems suggests that these elements were carried by molecular compounds accessible to gas-phase fractionation processes.

We now know the location of the primordial material in fractionation diagrams. Recently, the Genesis mission accurately measured the elemental composition of the Sun and, by extension, that of the solar protoplanetary disk \citep{Burnett03}. Over a period of three years, the solar wind was sampled by passive implantation of ions into pure target material. The spacecraft was located at Lagrangian point L1 to prevent any alteration of the solar wind composition by the terrestrial magnetic field. A result from Genesis of particular interest to this chapter was the measurement of the isotopic composition of solar oxygen, nitrogen and noble gases. Because the young Sun burned deuterium, the D/H ratio of the solar disk has been measured independently using the Jovian atmosphere. Genesis established that all solar system reservoirs (except for the atmosphere of Jupiter and probably of the other giant planets) are enriched, relative to the sun and solar protoplanetary disk, in the rare and heavy isotopes of hydrogen (D), nitrogen ($^{15}$N), oxygen ($^{17}$O and $^{18}$O). There are also indications of an enrichment in $^{13}$C in solar system solids and gases relative to the Sun \citep{Hashizume04}. 

\begin{figure}[ht!]
\begin{center}
\includegraphics[angle=0,scale=1.2]{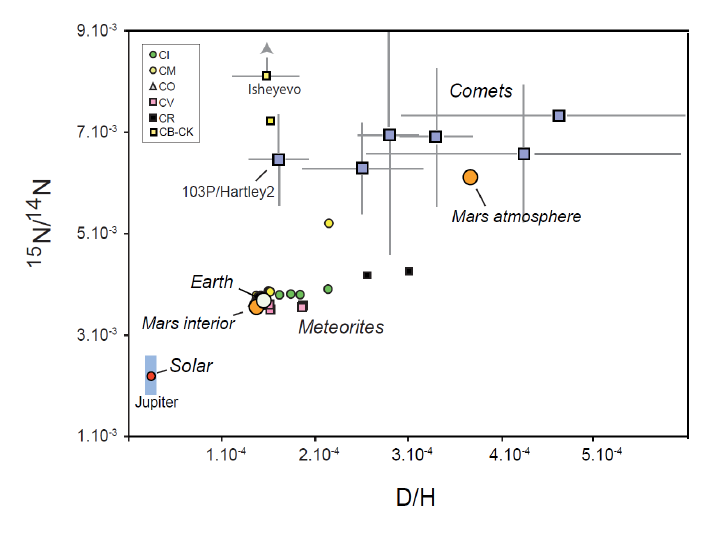}
\caption{Relation between the nitrogen ($^{15}$N/$^{14}$N) and deuterium (D/H) fractionation in the solar system.  Figure from \citet{Marty12}.}
\label{fig_iso}      
\end{center}
\end{figure}

The origin of the fractionation patterns in CNOH in the solar system is still a matter of vigorous debate, and, while data have improved in the past decade, few questions have been conclusively settled; we still do not have a clear idea of whether isotopic ratios in volatiles have an interstellar origin or whether they evolved as a consequence of protoplanetary disk evolution. In meteorites, the organic matter tends to show the strong isotopic anomalies of hydrogen and nitrogen \citep{Epstein87}. In the interstellar medium the $^{14}$N/$^{15}$N ratio is typically just above 400 with relatively little variation \citep[][and references therein]{Aleon10}. This is matched by the ratio in the Jovian atmosphere measured by the Galileo probe \citep[435$\pm$57,][]{Owen01}, and in the Sun \citep[442$\pm$66,][]{Marty10}, but contrasts with cometary and meteoritic values of $<$150 \citep[][and references therein]{Aleon10}. The D/H ratio in organics is similarly enhanced by factors of up to five in highly localized and heterogenous ``hot spots'' in meteorites, which are likely explained by the presence of organic radicals with D/H as high as $\sim$0.02 \citep{Remusat09}. The D/H ratio is generally correlated with the $^{14}$N/$^{15}$N fractionation in organics in meteorites and comets, relative to planetary atmospheres and the Sun (see Figure \ref{fig_iso}). Given the high levels of deuteration observed in the interstellar medium \citep{Ceccarelli07} and in protoplanetary disks \citep{Qi08}, it is tempting to use the meteoritic patterns to argue for an interstellar origin.  However, the deuteration in dense clouds and disks is typically higher than in meteorites, 
leading to disk models that use a highly fractionated initial condition that evolves toward a more equilibrated (less fractionated) system at the high temperatures and densities in a disk \citep[e.g.][]{Drouart99}. In this way, the protoplanetary disk phase may represent a modified ``intermediate'' fractionation state, relatively close to the solar D/H ratio, compared to the extreme primordial values. This is discussed in greater detail in the context of comets in \S\ref{sec:comets}.  We also refer the reader to the chapter in this volume focusing specfically on deuterium fractionation (Ceccarelli et al., this book).

Oxygen isotopes in the solar system uniquely display a puzzling mass-independent fractionation pattern \citep{Clayton73}, as shown in detail in Figure \ref{fig_O}.  The location of nearly all inner solar system materials on a single mass-independent fractionation line led to a search for physical processes capable of producing the observed fractionation slope, and affecting a large portion of the inner solar nebula.  A leading candidate is isotope-selective photodissociation of CO \citep{Kitamura83, Thiemens83, Clayton02, Yurimoto04, Lyons05}, a process that can now be investigated in protoplanetary disks, as we will discuss in Section \ref{sec:typical}.

\subsection{The origin of cometary volatiles in the solar system}
\label{sec:comets}
In the recent review of cometary chemistry by \cite{Mumma11}, the observed composition of comets is compared to that of the interstellar medium -- due to a lack of appropriate observations of protoplanetary disks. Using recent disk data, we can now begin to make a more direct comparison. The motivation for moving beyond the solar system is clear: Comets sample some of the oldest and most primitive material in the solar system, and are likely essentially unaltered since the protoplanetary disk stage. They are thus our best window into the volatile composition of the solar protoplanetary disk. A first, essential question to ask in preparation for a comparison to contemporary protoplanetary disks is: Is the volatile reservoir sampled by comets formed in the solar protoplanetary disk, or is it, instead, interstellar? 

The analysis of grains from the Stardust mission suggests that the answer is both, with captured samples of comet Wild 2 representing instead a wide range of origins.  The comet has incorporated some presolar material \citep{Brownlee06}, as evidenced by deuterium and $^{15}$N excesses in some grains \citep{Sandford06}, consistent with unprocessed materials in the outer solar system, where the comet is believed to have formed.  However, the cometary samples seem to most resemble samples of primitive meteorites, believed to have formed much closer to the young sun.  The oxygen, deuterium and nitrogen isotopic compositions of most Wild 2 grains are strikingly similar to carbonaceous chondrite meteorites \citep{McKeegan06}, and neon isotope ratios of noble gases are similar to those trapped in chondritic organics \citep{Marty08}. In addition, the grains are primarily crystalline, rather than amorphous, with compositions inconsistent with being formed via low-temperature condensation of interstellar grains, and therefore requiring a high temperature origin \citep{Brownlee06}.  Thus, Stardust suggests that cometary matter is a mixture of outer and inner solar system materials.

An alternative tracer of formation location may be the spin temperature of water. In molecules with two or more hydrogen atoms, transitions that change the total nuclear spin are strictly forbidden.  For water, therefore, as long as the chemical bonds are not broken, it is expected that the ratio between the ortho (s=0) and para (s=1) states persists at the value set at the formation temperature (spin temperature) of the molecule.  Thus, the ortho/para ratio in comets could potentially measure the formation location of the comet, and determine the relative contributions of in situ ice formation vs. inheritance of ice from the molecular cloud.  In comets, the water spin temperature has been measured to span from 26\,K \citep{Bockelee-Morvan09} to more than 50\,K \citep{Mumma88}. These temperatures are significantly higher than those of the water-forming regions in molecular clouds (10-15\,K), and perhaps represent a wide range of formation radii --- 5--50\,AU for typical protoplanetary disks \citep{Dalessio97,Chiang97}.  However, recent laboratory results complicate this picture considerably; two experiments show that the spin temperature of water produced at very low temperature and subsequently sublimated retains no history of its formation condition \citep{Sliter11, Hama11}.

Deuterium fractionation is another tracer of the cometary formation environment. The water ice in Oort cloud comets are typically enriched by a factor of two in deuterium (normalized to $^{1}$H) with respect to the terrestrial value and most primitive meteorites. However, the recent analysis of Jupiter comet 103P/Hartley2 \citep{Hartogh11} and Oort comet Garradd \citep{Bockelee-Morvan12} have shown that lower, even terrestrial-like, D/H ratios exist in comets. These observations were used to support the idea that Earth's water was delivered by comets -- a hypothesis that was otherwise falling out of favor based on the high D/H ratios in Oort comets. However, it also suggests that there is heterogeneity in the formation environments of comets, and that there may even be a gradient in the water D/H ratios in the solar disk, as predicted by models of the solar nebula \citep{Horner07}. This suggests a partial disk origin for cometary volatiles, or at least that a uniform interstellar D/H ratio evolved during the protoplanetary disk phase. 

The basic idea underlying this model is that the solar disk starts with an initial condition of highly deuterated water, inherited from the protosolar molecular cloud. The water then evolves,  via deuterium exchange with HD, toward less deuteration during the disk phase -- a process that happens faster at high temperatures in the inner disk, but not fast enough in the outer disk to account for the observed D/H ratios throughout the solar system (in chondrites as well as comets; \citealp{Drouart99}).  Consequently, models have invoked various mixing processes, such as turbulent diffusion or viscous expansion, to reproduce the observed deuteration from in-situ disk chemistry \citep{Mousis00, Hersant01, Willacy09, Jacquet13}. In particular, \citet{Yang12} find that Jupiter-family comets can obtain a terrestrial D/H ratio in the outermost part of a viscously-evolving disk. The final solar system D/H ratios are therefore likely dependent on {\it both} inheritance and in situ chemical evolution.

The cometary $^{14}$N/$^{15}$N ratios have been measured in both CN and HCN. In these species, the nitrogen isotopic ratio is a factor of 2 lower than that of the Earth and most meteorites, and a factor of 3 lower than that of the Sun. Enrichment of the heavy $^{15}$N isotope has been proposed to be evidence for interstellar fractionation, although chemical models of dense clouds do not readily produce fractionation in any nitrogen carrier as strong as that seen in the solar system \citep{Terzieva00}, and the usability of the nitrogen isotopic ratio as an origin tracer remains unclear. 

A final complication in the story of comets is that it is now understood that the orbital parameters of a given comet are a poor tracer of its formation radius in the solar disk. The Nice model \citep{Gomes05} suggests that comets forming anywhere from 5 to $>$30 AU may have contributed to both the Kuiper Belt comets as well as the Oort cloud comets. Cometary origins may be even more interesting if, as suggested by \citet{Levison10}, $\sim$50\% of comets are inherited from other stellar systems.

\subsection{The total mass of condensible volatile molecules}
\label{IceMass}
The magnitude of the local solid surface density is linked to the efficiency of planetesimal formation, and by extension the formation of planets via accretionary processes. \cite{Hayashi81} originally used an ice/rock ratio of 4 in his formulation of the classical minimum-mass solar nebula (MMSN). Some investigations in later years have revised this value slightly downwards, but it is still high. \citet{Stevenson85} found a value of 2--3, based on comparisons to the icy moons of the solar system giant planets. \cite{Lodders03} found ice/rock = 2, based on equilibrium chemistry. The recent update to the MMSN by \cite{Desch07} also uses ice/rock = 2, while \cite{Dodson-Robinson09} used a full chemical model coupled with a dynamical disk model to predict an ice/rock ratio close to the original ice/rock=4. This variation is in part due to differences in assumptions on the elemental oxygen abundance, and on the chemistry. The higher values for the ice/rock ratio require that both the bulk carbon and nitrogen are readily condensible with sublimation temperatures higher than those of e.g., CO and N$_2$. Typically, this requires carriers such as CH$_4$ and NH$_3$. A high value for the ice/rock ratio therefore predicts large amount of gaseous CH$_4$ and NH$_3$ just inside their respective snow lines. The theoretical expectation for protoplanetary disks is therefore that the value most likely lies somewhere between ice/rock$\sim$2--4, and precise observations of all the major carriers of carbon, oxygen and nitrogen are needed to determine the true ice/rock ratios, as well as their potential variation from disk to disk.

\begin{figure}[ht!]
\includegraphics[width=8cm]{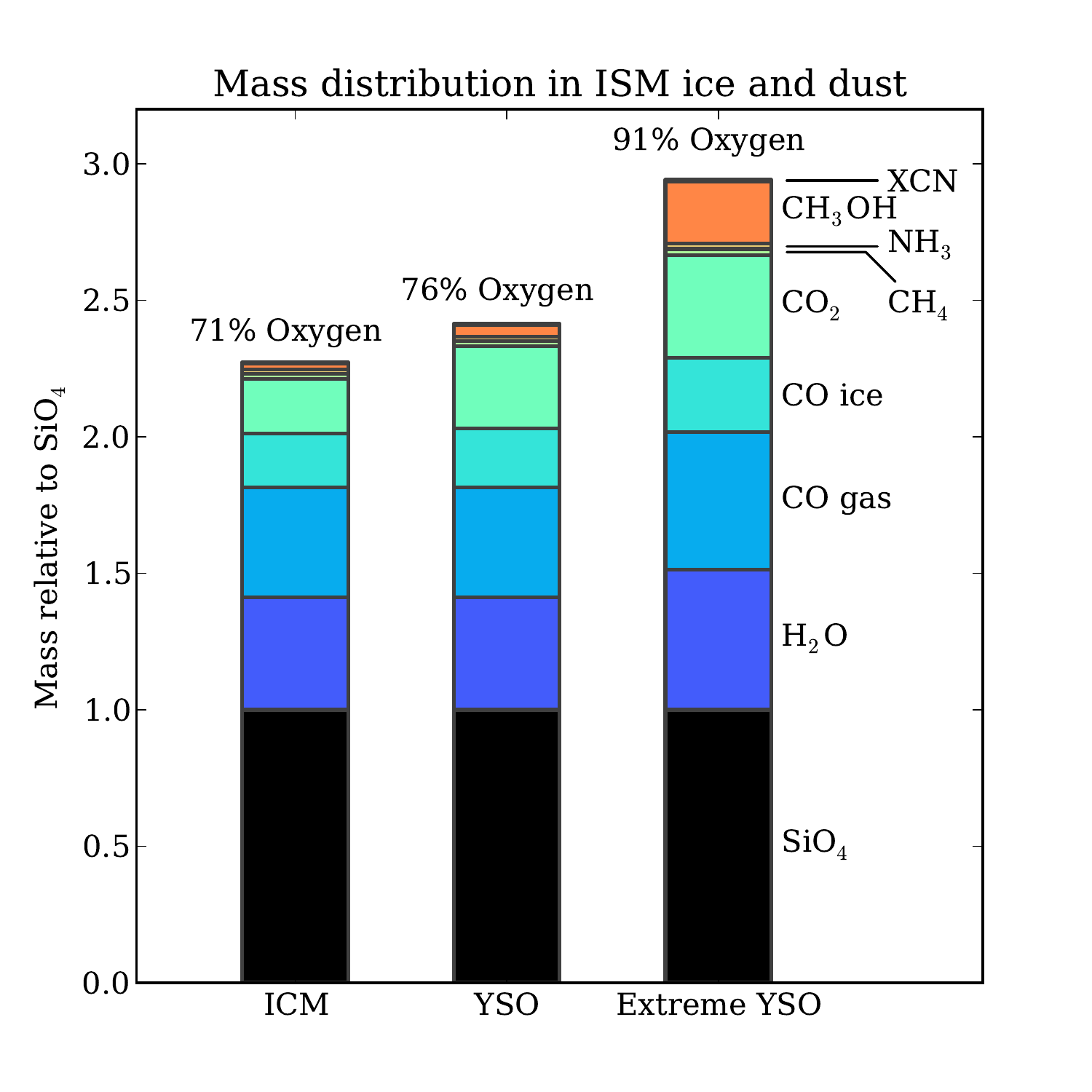}
\caption{The mass distribution of molecules in the interstellar medium derived from ice observations \citep{Whittet07,Boogert08,Oberg11}. The distribution assumes that all available silicon is bound in the form of tetrahedral silicates, such as forsterite (taking up four oxygen atoms per silicon atom), although other common silicate species contain slightly less oxygen (three oxygen atoms per silicon atom), including enstatite and forsterite. The plot displays only observed carriers (except for silicates) for three environments: The intra-cloud medium (ICM), young stellar objects and the ``extreme'' YSO with the highest observed ice abundances from \cite{Pontoppidan04}. Above each column is shown the fraction of the solar elemental oxygen abundance that is included in each budget. The observed oxygen fractions are slightly higher than in \cite{Whittet10} due to our use of the lower \cite{Grevesse10} value for $\rm O/H=4.89\times 10^{-4}$, instead of the $\rm O/H=5.75\times 10^{-4}$ value of \cite{Przybilla08}. } 
\label{fig:ISM_budget}
\end{figure}

In the dense interstellar medium and in protostellar envelopes, a nearly complete inventory of ice and dust has been observed and quantified \citep[e.g.,][]{Gibb04,Whittet09,Oberg11} using infrared absorption spectroscopy toward young stellar objects and background stars. \cite{Whittet10} finds that as much as half of the oxygen is unaccounted for in dense clouds when taking into account refractories -- such as silicates, as well as volatiles in the form of water, CO$_2$ ices and CO gas. This potentially lowers the ice/rock ratio if the missing oxygen is in the form of a chondritic-like refractory organic phase (as suggested by Whittet) or as extremely volatile O$_2$. In the densest regions of dark clouds and protostellar envelopes, it has been suggested that the ice abundances increase, in particular those of H$_2$O, CO$_2$ and CH$_3$OH, and some observations find that as much as 90\% of the oxygen may be accounted for in one region (\citealp{Pontoppidan04}; see also Figure \ref{fig:ISM_budget})). In the case where the ice abundances are the highest, this yields an observed ice/rock ratio of at least $\sim 1.5$. Strictly speaking, this is a lower limit, since some ice species are as yet unidentified \citep{Boogert08}, and some species, such as N$_2$ \citep{Pontoppidan03}, may not be detectable. Importantly, in the ISM, NH$_3$ is not very abundant \citep[$\sim 5\%$ relative to water,][]{Boogert08}. The remaining uncertain components are the main reservoirs of carbon and nitrogen, with detected reservoirs of the former only accounting for 35\% of the solar abundance, and detected reservoirs of the latter accounting for less than 10\% of the solar abundance. Likely reservoirs in the ISM are refractory carbon and N$_2$, neither of which is directly identifiable using infrared spectroscopy. The difference up to the putative solar value of ice/rock = 4 is essentially recovered if the missing oxygen, carbon and nitrogen are fully sequestered into ices; in \cite{Dodson-Robinson09} for example, 50\% of the C is in the form of CH$_4$, and 80\% of N in the form of NH$_3$.  Thus, whether the bulk disk chemistry resembles the ISM or not is of fundamental importance for computing the disk's solid surface density.  Recent observations have begun to reveal some of the chemical demographics of protoplanetary disks, as we shall see in the following section.

\section{\textbf{The distribution and evolution of volatiles in protoplanetary disks}}
\label{sec:evolution}
Protoplanetary disks are structures characterized by a wide range of physical environments. They contain regions with densities in excess of $\rm 10^{15}\,cm^{-3}$ and ISM-like conditions with densities $\rm <10^4\,cm^{-3}$, they have temperatures in the range $10\,$K$\ <T<\ $10,000$\,$K, and parts of them are deeply hidden beneath optically thick layers of dust with depths of more than 1,000 magnitudes. From an astrophysical perspective, disk conditions are often characterized in radial or vertical terms, depending on circumstances, although of course any disk has a full three-dimensional structure. Radially, disks are often divided into vaguely defined regions that are roughly separated by an assumed mode of planet formation: Terrestrial (0-5\,AU), giant planet (5-20\,AU) and Kuiper/cometary belt ($>$20\,AU). The first two are characterized by a series of condensation fronts, and are often bundled into an ``inner disk'' nomenclature. The latter region is universally known as the ``outer disk''. Vertically, disks are thought of as consisting of three regions --- a low-density, molecule-poor region near the disk surface, an intermediate region, with sufficient radiation shielding and temperatures to sustain and excite gas-phase molecules (often called the ``warm molecular layer''), and a colder region near the disk midplane which is often shielded by optically thick dust and in which volatile molecules may condense beyond a certain radius. A more general review of protoplanetary disk structure as it relates to chemistry can be found in \cite{Henning13}.

The methods used to infer the presence and relative abundances of disk volatiles draw from a heterogenous set of observational facilities, spanning the ultraviolet to radio regimes. Advances in observational capabilities across the electromagnetic spectrum have allowed more comprehensive descriptions of the bulk chemical composition of protoplanetary disks than previously possible. In place of a picture in which disks are composed simplistically of molecular hydrogen and helium with trace amounts of refractory dust, they are now seen as complex environments with distinct regions characterized by fundamentally different chemical compositions. Some of these regions are more accessible to observations than others, and the more readily observed regions are often used to infer the properties of unseen or hidden regions. With a few notable exceptions of disks with very high accretion rates, protoplanetary disks are generally characterized by having a surface that is hotter than the interior. This property leads to observables (line as well as continuum emission) that are dominated by the surface of the molecular layer. This tends to be true even at radii and at observing wavelengths where the disks are optically thin in the vertical direction \citep{Semenov08}. Thus, most of our direct knowledge of volatiles in disks pertains to surface conditions and to a fraction of the vertical column density.

\vspace{0.15cm}
\subsection{History of observations of volatiles in disks}
\vspace{0.15cm}

Until a few years ago, direct measurements of the amount and distribution of bulk volatiles in protoplanetary disks were limited. Part of the reason was that the strong transitions of the most common volatile species were not visible from the ground and space-based observatories were still too insensitive to obtain spectroscopy of typical protoplanetary disks. Millimeter observations of a small number of large, massive disks revealed the presence of trace species ($\lesssim 1\%$ of the total volatile mass) in the cold, outer regions of disks (beyond 100\,AU), such as CS, HCO$^+$ and HCN \citep{Dutrey97,Thi04,Oberg10}. An exception to these millimeter-detected trace species is CO, which is one of the most important bulk volatiles. A key conclusion of the early millimeter work was that efficient freeze-out leads to strong depletions in the outer disk \citep[e.g.,][]{Qi04}. This was supported by a few detections of ices in disks. \cite{Malfait98} and \citet{Chiang01} discovered emission from crystalline water ice in two disks in the far-infrared. At shorter wavelengths, ices do not produce emission features since the temperatures required to excite existing resonances below $\sim$40\,$\mu$m will effectively desorb all bulk volatile species, including water. Scattered near-infrared light has since revealed water ice in at least one disk \citep{Honda09}, and attempts have been made to measure ices in absorption toward edge-on disks {\citep{Thi02}}, although difficulties in locating the absorbing material along the line of sight has led to ambiguities between ice located in the disk itself and ice located in foreground clouds \citep{Pontoppidan05}.  More recently, sensitive searches for ices in edge-on disks have made progress in detecting ice absorption in edge-on disks with no potential contribution from cold foreground cloud material \citep{Aikawa12}.  In addition, \citet{McClure12} find a tentative detection of water ice emission from a disk observed with the Herschel Space Observatory, and recent upgrades to PACS data reduction routines may enable additional detections in the near future.

Exploration of the molecular content of warm inner disks ($\lesssim 10\,$AU) has greatly expanded since PPV. Inner disks have been readily detected in CO ro-vibrational lines \citep{Najita03,Blake04}, where it is essentially ubiquitous in protoplanetary disks \citep{Brown13}. \cite{Carr04} detected lines from hot water near 2.5\,$\mu$m toward one young star and argued that the origin was in a Keplerian disk.   However, it was with the launch of the Spitzer Space Telescope in August 2003 that this field underwent a major advance in understanding.  C$_2$H$_2$, HCN and CO$_2$ were first detected by Spitzer in absorption toward a young low-mass star \citep{Lahuis06}, with C$_2$H$_2$, HCN absorption also being detected from the ground in a second source \citep{Gibb07}.  But Spitzer also initiated the very first comprehensive surveys of the molecular content in the inner regions of protoplanetary disks.  \cite{Carr08} realized that the mid-infrared spectrum from 10-20\,$\mu$m of the classical T Tauri star AA Tau showed a low-level complex forest of molecular emission, mostly from pure rotational lines of water, along with lines and bands from OH, HCN, C$_2$H$_2$ and CO$_2$. The emission was characteristic of abundant water with temperatures of 500-1000\,K. \cite{Salyk08} discovered an additional two disks with strong mid-infrared molecular emission, and showed that these disks also have emission from hot water as traced by ro-vibrational transitions near 3\,$\mu$m. A survey of more than 60 protoplanetary disks around low-mass ($\lesssim 1\,M_{\odot}$) stars revealed the presence of mid-infrared emission bands from organics (HCN in up to 30\% of the disks, and up to 10\% for C$_2$H$_2$) \citep{Pascucci09}. Advances in the data reduction procedure for high resolution Spitzer spectra subsequently revealed water emission, along with emission from the organics, OH, and CO$_2$, in about half the sources in a sample of $\sim 50$ disks around low-mass stars \citep{Pontoppidan10}. Following the initial flurry of results based on Spitzer data, the later years have focused on modeling and interpreting these data \citep{Salyk11,Carr11}, as well as obtaining follow-up observations in the near- and mid-infrared of e.g., water and OH, by ground-based facilities at high spectral resolution \citep{Pontoppidan10b,Mandell12}.

\vspace{0.15cm}
\subsection{Retrieval of volatile abundances from protoplanetary disks}
\vspace{0.15cm}
Before proceeding to the implications of the observations of volatiles in protoplanetary disks, it is worthwhile to briefly consider the technical difficulties in accurately measuring the basic physical and chemical parameters of disks from observations of molecular emission lines. The assumptions that are imposed on the data analysis essentially always generate model dependencies that can cloud interpretation. For this reason intense work has been underway for the past decade to minimize the model dependencies in this process and to better understand the physics that forms emission lines from disks, as well as other analogous nebular objects. {\it Retrieval} refers to the modeling process of inverting line fluxes and profiles to fundamental physical parameters such as the fractional abundance $(X({\rm species}) = n({\rm species})/n({\rm H})(R,z)$), the total gas mass or kinetic gas temperature. 

Because molecular emission line strengths are dependent on abundance as well as excitation, retrieval is necessarily a model-dependent, usually iterative, process, even for disks that are spatially resolved. A common procedure uses broad-band multi-wavelength observations (1-1000\,$\mu$m) to measure the emission from dust in the disk. Structural models for the distribution of dust in the disk are fit to the data, typically under some assumption of dust properties. Assuming that the dust traces the bulk molecular gas, the spatial fractional abundance structure of molecular gas-phase species can be retrieved by modeling the line emission implied by the physical structure fixed by the dust \citep{Zhang13,Bergin13}. Another analysis approach does not attempt to retrieve molecular abundances directly, but attempts to constrain the global disk physical structure by fixing the input elemental abundances and using a thermo-chemical model to calculate line fluxes. Such models simultaneously solve the chemistry and detailed radiative transfer, but rely on having incorporated all important physical and chemical processes \citep[e.g., ][]{Aikawa02,Thi10,Woitke11,Tilling12}. Both approaches carry with them significant uncertainties and degeneracies, so molecular abundances from disks are probably still inaccurate to an order of magnitude, although progress is continuously being made to reduce this uncertainty \citep{Kamp13}.

\vspace{0.15cm}
\subsection{Demographics of volatiles in disks}
\vspace{0.15cm}

\begin{figure}[ht!]
\centering
\includegraphics[width=8cm]{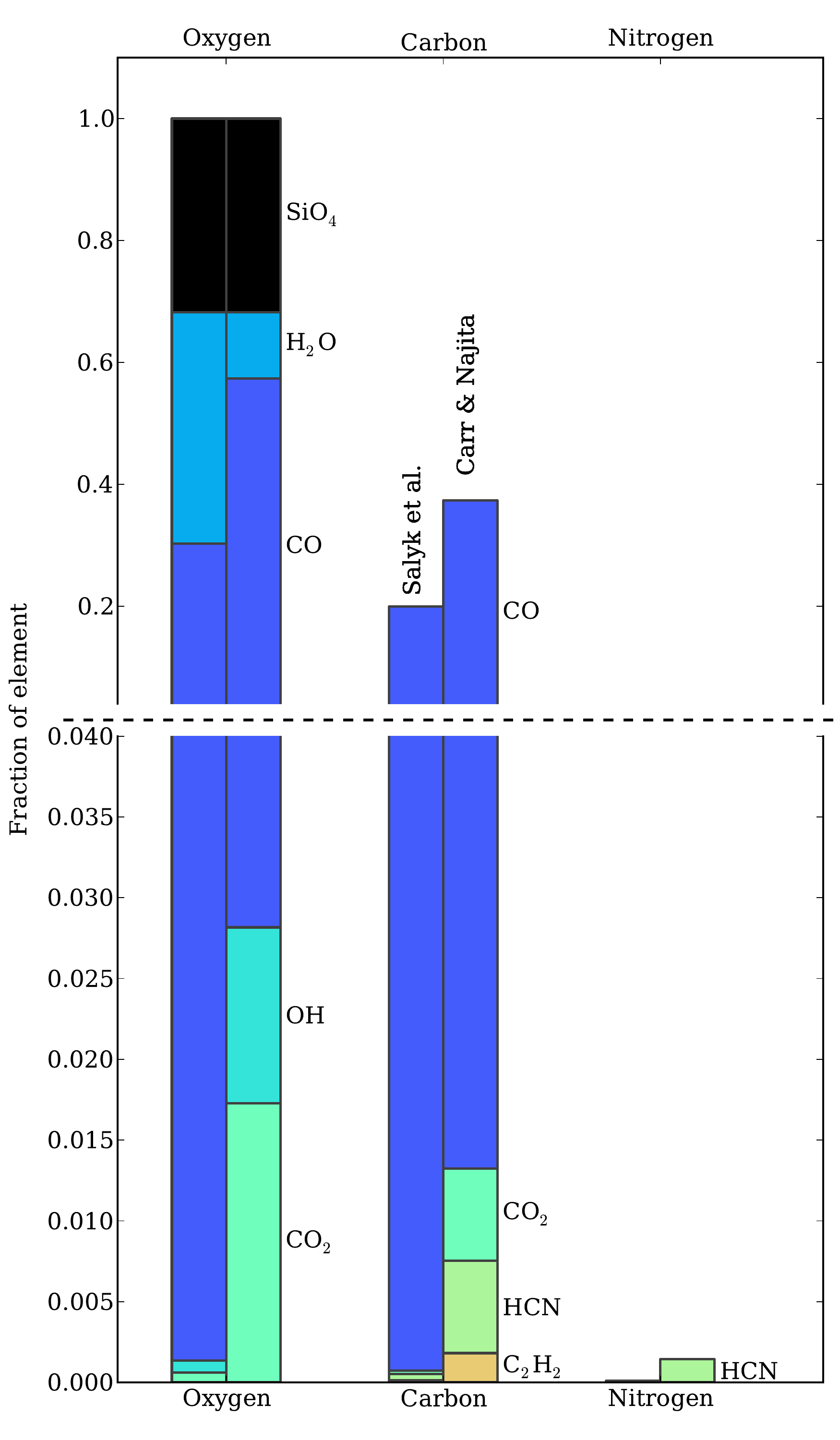}
\caption{Observed elemental abundance budget at $\sim$1 AU in protoplanetary disks. The values retrieved by \cite{Salyk11}, \cite{Carr08} and \cite{Carr11} are compared side-by-side to provide a rough indication of current uncertainties. All budgets are normalized to the assumption that all oxygen is accounted for in the warm molecular layer (fraction of O$=1$). With this assumption, about a 3rd of the carbon is accounted for, and almost none of the nitrogen. The implication is that any missing contribution is located in an as-yet unobserved carrier. Note that the plot has been split into two different y-axis scales to simultaneously show major as well as minor carriers. }
\label{fig:disks_budget}
\end{figure}

It is now possible to summarize some general observed properties of volatiles in protoplanetary disks. Their composition is dominated by the classical carbon-nitrogen-oxygen elemental trifecta as shown in Figure \ref{fig:disks_budget}. In addition there are likely trace species including other nucleosynthetic elements (e.g., sulfur and fluorine), although the only firm detections to date are of sulfur-bearing species, and those appear to indicate that the bulk of the sulfur is sequestered in a refractory carrier \citep{Dutrey11}. 

Comparing the observed column densities of CNO-bearing volatiles in the inner disks, \cite{Salyk11} find that, while the data are consistent with most of the oxygen being carried by water and CO, carbon seems to be slightly under-abundant by a factor of 2--3, while nitrogen (only observed in the form of HCN) is still largely unaccounted for by about 2 orders of magnitude. \cite{Carr11} used a marginally different analysis method, but found slightly higher HCN and C$_2$H$_2$ abundances by factors of a few in similar, albeit not identical, disks. This, while underlining that molecular abundances retrieved from low-resolution spectra are still uncertain, does not change the basic demographic picture. The (slight) carbon deficit may be explained by some sequestration in a carbonaceous refractory component, similar to the situation in the ISM. This is, however, not consistent with the carbon deficit in the inner solar system, as discussed in \S\ref{carbon_def}. The nitrogen deficit is potentially interesting as NH$_3$ can be ruled out as a significant carrier in the inner disk by stringent upper limites on the presence of emission bands at 10\,$\mu$m \citep{Salyk11} and 3\,$\mu$m \citep{Mandell12}. This has led to speculation that most of the nitrogen in the terrestrial region is stored in N$_2$, and is largely unavailable for planetesimal formation. Beyond the terrestrial region toward the outer disk, we would still expect to find nitrogen in the form of NH$_3$ to explain the cometary and giant planet abundances. The inner disk elemental abundances are summarized in Table \ref{AbundanceTable}, and illustrated in Figure \ref{fig:disks_budget}. 

\begin{table}[ht!]
\caption{Distribution of C, N and O in inner disks.\tablenotemark{a}}
\begin{tabular}{llll}
\hline
\hline
&Oxygen & Carbon & Nitrogen\\
&\% & \% & \%\\
\hline
H$_2$O      &38   &--&-- \\
CO             &30   &57&-- \\
HCN          &--   &0.1-0.6\tablenotemark{b}&0.4-2.5 \\
NH$_3$     &--   &--&$<1.1$\\
C$_2$H$_2$   &--   &0.03-0.25&-- \\
CO$_2$     &(0.04)&(0.05)&-- \\
OH            &(0.04)&--& -- \\
Silicates    &32   &--&-- \\
\hline
Total         &100\tablenotemark{c}&57-58&0.4-2.5 \\
\end{tabular}
\tablenotetext{a}{Values are relevant for disks around stars of roughly 0.5-1.5\,$M_{\odot}$.}
\tablenotetext{b}{Ranges indicate differences in median derived values from \\ \cite{Salyk11} and \cite{Carr11}.}
\tablenotetext{c}{It is assumed that all the oxygen is accounted for, and the other \\fractions are scaled using the solar elemental abundances.}
\label{AbundanceTable}
\end{table}

As a comparison to the elemental budgets determined from molecular lines, a number of optical integral-field spectroscopic measurements of the abundances of many elements in the photoevaporative flows from protoplanetary disks in Orion are available \citep{Tsamis11,Mesa-Delgado12,Tsamis13}. The Orion disks are being eroded by the harsh interstellar radiation field from young massive stars, providing a unique opportunity to measure their bulk composition. They show almost no depletion. In the cases of carbon, oxygen, the abundances are within a factor 1.5 relative to solar, while the nitrogen abundance is a little more than a factor 2 lower than solar. This suggests that most of the unseen molecular carriers are being entrained in the photoevaporative flows. 

In the outer disk, the CNO budget is observationally still incomplete due to strong depletions from the gas-phase by freeze-out. In general most molecules are depleted relative to cloud abundances \citep{Dutrey97} and CO is underabundant by about an order of magnitude \citep{vanZadelhoff01, Qi04}. Water is depleted by up to 6--7 orders of magnitude \citep{Bergin10,Hogerheijde11} - possibly even more than can be explained by pure freeze-out.  

\vspace{0.15cm}
\subsection{Chemical dependencies on stellar type}
\vspace{0.15cm}

\begin{figure}[ht]
\includegraphics[width=8cm]{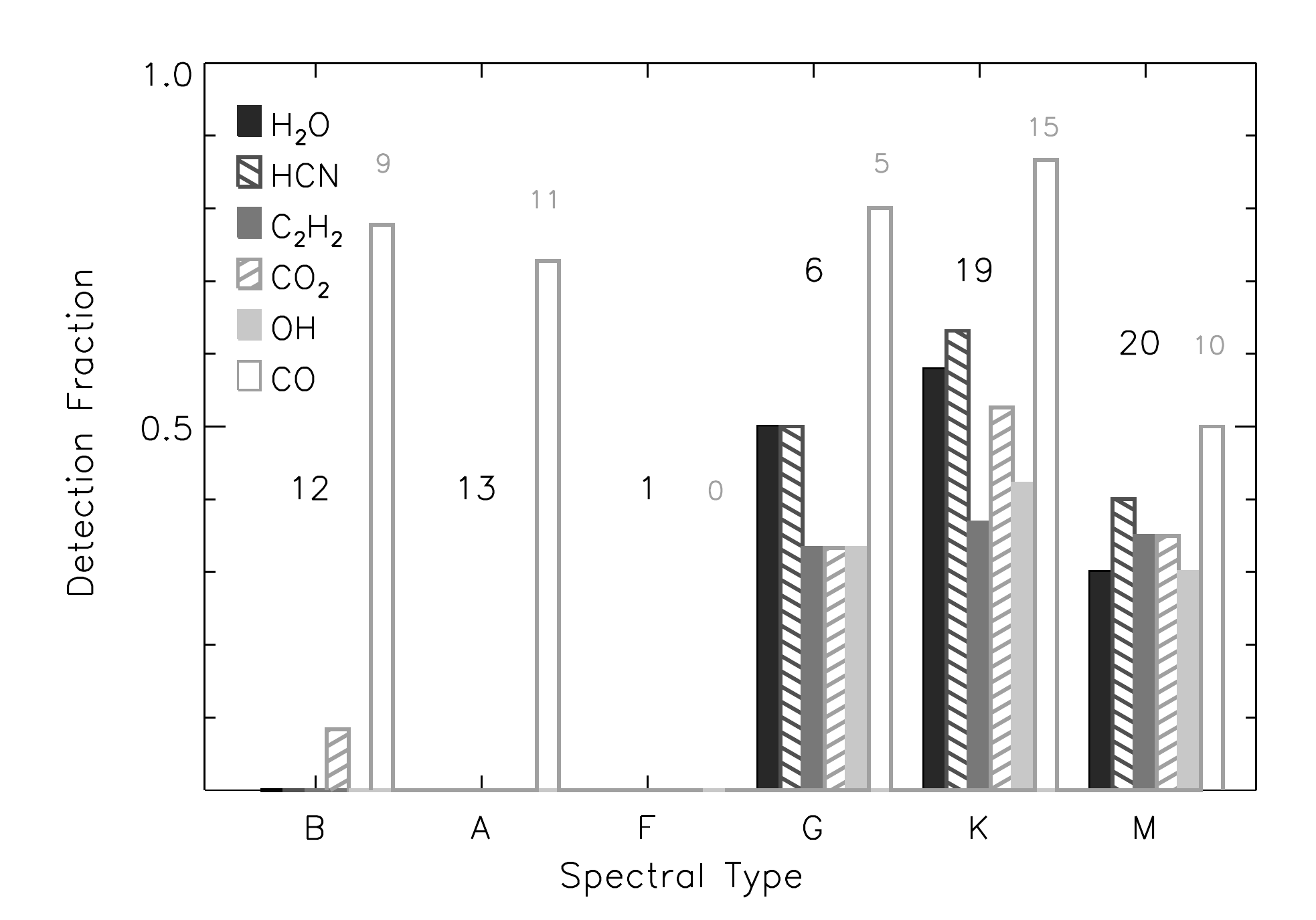}
\caption{The relation between Spitzer detection rates of the infrared molecular forest and stellar spectral type \citep[from][]{Pontoppidan10}.}
\label{fig:herbig_rate}
\end{figure}

There are large differences in the observational signatures of volatiles as a function of the mass and evolutionary stage of the central star. It is generally difficult to detect molecular emission from disks around Herbig Ae stars (see Figure\ \ref{fig:herbig_rate}), and the implications of this strong observable are currently being debated. The effect is strongest in the inner disk, where very few disks around stars hotter than $\sim 7000\,$K show any molecular emission at all \citep{Pontoppidan10}, with the exception of CO \citep{Blake04} and OH \citep{Mandell08,Fedele11}. The apparent increases of the OH/H$_2$O ratio in Herbig stars have led to speculation that the strong dependence of molecular line detections on spectral type is due to the photodestruction of water in the photolayers of Herbig disks, rather than an actual deficit of bulk midplane water. However, it is possible that an enhanced abundance of the photo-products of water, and radicals in general, in combination with efficient vertical mixing, can lead to large changes in bulk chemistry dependent on spectral type. It should be noted, however, that it is still debated how much of a depletion the non-detections actually correspond to, as the dynamic range of the infrared Spitzer spectra is relatively low. A detection of water in the disk around HD 163296 suggests that the depletion in the inner disk in some cases may be as little as 1--2 orders of magnitude, although in this case, the water emission originates from somewhat larger radii (15-20\,AU) than traced by the multitude of mid-infrared molecular lines \citep[$\sim$1\,AU, ][]{Fedele12}.

Toward the lower end of the stellar mass spectrum, \citet{Pascucci09} find that disks around young stars with spectral types M5--M9 have significantly less HCN relative to C$_2$H$_2$, possibly requiring a greater fraction of the elemental nitrogen being sequestered as highly volatile N$_2$. Further emphasizing the dependence of stellar type on inner disk chemistry, \cite{Pascucci13} recently found strongly enhanced carbon chemistry in brown dwarf disks, as measured by HCN and C$_2$H$_2$, relative to oxygen chemistry, as measured by H$_2$O. Observationally, solar-mass young stars therefore seem to be the most water-rich, and research in the near future is likely to explore how this may affect the structure and composition of planetary systems around different stars. 

\vspace{0.15cm}
\subsection{Origin of disk volatiles: Inheritance and reset}
\vspace{0.15cm}

How much of the proto-stellar volatile material survives incorporation into the disk? That is, are initial disk conditions predominantly protostellar \citep{Visser09} and to what degree do the initial chemical conditions survive the physical evolution of protoplanetary disks? Which regions of the disk are {\it inherited} giving rise to planetesimals of circum- or interstellar origin, and which regions are chemically {\it reset}, leading to disk-like planetesimals? Chemical models indicate that essentially all disk regions are chemically active on time scales shorter than the lifetime of the disk. The dominant chemical pathways and the timescales for equilibriation, however, are expected to vary substantially from region to region, depending on the physical properties of the local environment \citep{Semenov11}. Partial answers to these questions may be found in the observed volatile demographics and their comparison to protostellar chemistry. Different scenarios should result in significant differences in volatile composition and isotope fractionation. Searching for such differences or similarities is the subject of broad ongoing observational investigations. 

\begin{figure}[ht]
\includegraphics[width=8cm]{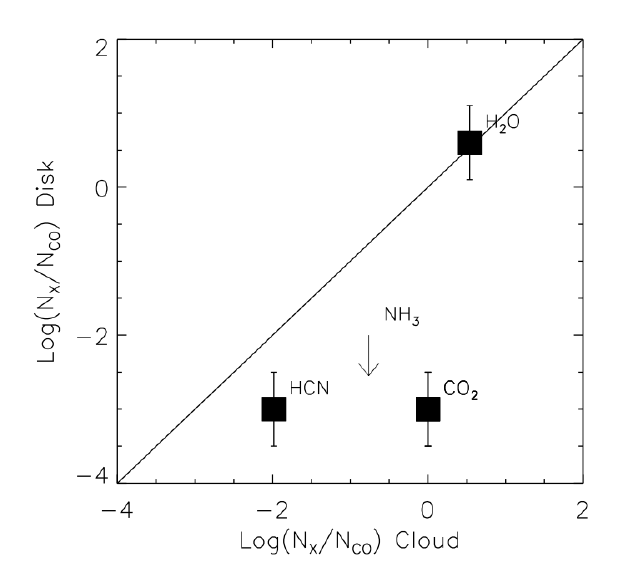}
\caption{Comparison between the observed abundances of gas-phase inner disk volatiles derived from Spitzer-IRS spectra \citep{Salyk11} relative to those in ices in protostellar clouds \citep{Oberg11,Lahuis00}.  Disk abundances are appropriate for the inner disk, as the Spitzer-IRS emission lines originate primarily in the few AU region.}
\label{fig:cloud_disk}
\end{figure}

Prestellar and protostellar volatile compositions have been systematically investigated through gas and ice observations \citep{vanDishoeck04,Oberg11,Caselli12}. There is some familial resemblance between the protostellar and comet compositions in our Solar System, which has been taken as evidence of a direct inheritance of the comet-forming volatile reservoir \citep{Mumma11}. The possibility of such a connection is supported by recent two-dimensional hydrodynamical simulations of the chemical evolution on trajectories from protostellar envelopes up to incorporation into the disks \citep{Aikawa99,Brinch08,vanWeeren09,Visser09,Visser11,Hincelin13}. While different trajectories are exposed to different levels of heat and radiation, most ices that are deposited into the comet-forming region survive incorporation, and only some of the most volatile ices sublimate during disk formation (e.g. N$_2$ and CO). 

Some ice evolution during disk formation is likely, however, since the grains spend a significant time at 20-40\,K -- temperatures known to be favorable for complex molecule formation \citep{Garrod06,Garrod08,Oberg09}. Once the disk is formed, the initial volatile abundances may evolve through gas and grain surface reactions. The importance of this {\it in situ} processing compared to the chemistry preceding the disk stage will depend on the exposure of disk material to conditions that favor chemical reactions, i.e. UV and X-ray radiation, high densities and heat. Disk models that begin with mostly atomic structures implicitly assume that the disk chemistry is reset. This assumption is plausible for regions inside the snow line because of the short chemical timescales ($10^4$ years or less) for regions dominated by gas-phase chemistry \citep{Woitke09}. This is consistent with the meteoritic evidence that has always favored a complete chemical reset; most of the collected meteorites originated from feeding zones within the solar snow line (see also Section \ref{SolarNebula} for a discussion of the evidence for a reset from a cosmochemical perspective). 

The addition of accretion flows and advection, which transport volatiles from the outer to the inner disk, as well as vertical and radial mixing driven by local turbulence, result in mixing between a chemically active surface and a relatively dormant midplane.  \citet{Semenov11} explored the different turbulence and chemistry timescales in disks, deriving characteric timescales to determine which chemical processes can be efficiently erased by mixing (with necessary assumptions about disk parameters, including ionization rates, structure, etc.).  They found that vertical mixing is faster than the disk lifetime ($<10^6$ years) interior of 200 AU in a typical disk. Ion-molecule reaction rates depend on the abundances of ions involved in the reaction, but can be fast; as a representative example, they estimate rates leading to the production of HCO$^+$ to be $<10^4$ years throughout the disk.  In such a case, the gas-phase chemistry will not retain any memory of the original composition, and inside the snow line, it is very difficult to retain any primordial signature of volatiles over the disk life time. Photochemistry is fast outside of the midplane. Outside the snowline, the midplane is dominated by fast freeze-out and a slower desorption by cosmic rays that still has a timescale shorter than the disk lifetime, implying that some gas-grain circulation will take place. Grain surface chemistry can in some cases be very slow, however, and thus icy grains that do not get dredged up into the disk atmosphere or into the inner disk may retain an inherited chemistry. Gas-phase emission from disks should mainly reflect disk {\it in situ} chemistry, especially from regions inside the main condensation fronts, including e.g. isotopic fractionation patterns, while comet compositions and probes of disk ices from beyond a fews 10's of AU should exhibit protostellar-like abundances and fractionation patterns.

{\it In situ} deuterium fractionation in disks provides the most obvious explanation for the recent observations of different disk distributions of two deuterated molecules, DCN and DCO$^+$, with formation pathways that present different temperature dependencies \citep{Oberg12}. As discussed in Section \ref{sec:fractionation}, cometary H$_2$O deuterium fractionation may or may not be consistent with protostellar values, while nitrogen fractionation as well as NH$_3$ spin temperatures both point to single cold origin \citep{Mumma11}, which would require that part of the disk (most likely the disk midplane) retain a history of the initial (cloud) conditions.

So how do the chemical abundances of solar-type inner disks compare to those of protostellar clouds and envelopes?  Figure \ref{fig:cloud_disk} shows a comparison of the abundances of HCN, CO$_2$, H$_2$O and NH$_3$ (relative to CO) derived from Spitzer-IRS observations of disks \citep{Salyk11} with those derived from protostellar clouds \citep{Oberg11}.  While the abundance of H$_2$O is similar for both clouds and disks, inner disk abundances of other observed species are lower than in clouds. In summary, evidence for significant changes in inner disk chemistry during disk formation and evolution is emerging, emphasizing that protoplanetary disks have inner regions strongly affected by a reset-type chemistry. To what degree this is consistent with the cosmochemical record will be an active research subject in the near future. 

\vspace{0.15cm}
\subsection{Volatiles during transient accretion events}
\vspace{0.15cm}
  
Chemical models of disk evolution often assume a relatively quiescent disk, experiencing processes no more dynamic than viscous accretion. However, evidence is mounting that accretion of material from the natal cloud to the disk must occur in short, intense bursts that may drive the chemical reset over a relatively wide range of disk radii (\citealp[e.g.][]{Dunham10}; see also the chapter on young eruptive variables elsewhere in this book). The outbursts of FU Ori and EXor stars reveal violent and energetic accretion episodes \citep[e.g.][]{Herbig66, Herbig77, Hartmann96} that must have profound effects on the volatile contents of their disks. Not surprisingly, the molecular emission lines from eruptive variables show remarkable variability throughout outbursting events. For example, observations of ro-vibrational CO in the EXor V1647~Ori revealed fundamental and overtone CO emission at the beginning of its outburst \citep{Vacca04, Rettig05}, cooler emission and a blue-shifted absorption after return to pre-outburst luminosity, and finally a ceasing of all emission after one year \citep{Brittain07}.  Variable rovibrational lines of CO have also been observed towards EX Lupi, show evidence for at least two gas components at different temperatures, with one tightly correlated with the system luminosity, and displaying asymmetries due to a possible disk hotspot, and evidence for a warm molecular outflow \citep{Goto11}. Recently, a number of additional molecular species have been observed in EX Lupi both before and during an outburst taking place during 2008, viz. H$_2$O, OH, C$_2$H$_2$,  HCN and CO$_2$ \citep{Banzatti12}. When the star underwent the outburst, the organic lines went away while the water and OH lines brightened, suggestive of the simultaneous evaporation of large amounts of water locked in ice pre-burst, and active photo-chemistry in the disk surface.  

\subsection{Was the solar nebula chemically typical?}
\label{sec:typical}
Although destined to grow dramatically in the future, comparisons of protoplanetary disk chemistry with aspects of the cosmochemical evidence from the solar nebula are emerging. It indeed appears that critical structures in the solar nebula with respect to volatiles are reproduced in protoplanetary disks. The astrophysical perspective shows a clear distribution of a volatile-rich inner nebula and volatile-poor outer nebula, with different snowlines between water and more volatile CO. 
   
For instance, do the D/H fractionation patterns in protoplanetary disks match those of the solar system? While still elusive in disks, HDO and H$_2^{18}$O have been detected in the warm envelopes in the immediate surroundings of young solar mass stars. The measurements have a wider degree of diversity than observed in solar system materials, with ratios of warm HDO to H$_2$O ranging from $ < 6 \times 10^{-4}$ \citep[][]{jvd10} to as high as $6-8 \times 10^{-2}$ (\citealp{Coutens12,Taquet13}; corresponding to D/H ratios a factor of two lower). Part of the explanation for this heterogeneity could be that the derivation of the D/H ratio depends sensitively on whether the observations resolve the innermost (presumably disk-like) regions \citep{pjvd13}, as well as on the details of the retrieval method, with recent models deriving lower values of D/H \citep{Visser13, Persson13}. Nevertheless, the lower values of D/H for many sources are similar to those of cometary ices. We direct the reader to the chapter by van Dishoeck et al. for more details about the analysis and interpretation of water vapor observations in young, embedded disks.

Protoplanetary disks also show tentative evidence in support of proposed solutions to the carbon deficit problem described in Section \ref{carbon_def}.  [C I] has not yet been detected in disks, with strong upper limits suggesting a general depletion of gas-phase carbon relative to the observed dust mass, even after accounting for CO freeze-out \citep{Panic10, Chapillon10, Casassus13a}.    While a possible explanation for this depletion is that the overall gas/dust ratio of the disk is depleted, \citet{Bruderer12} show that, for at least one disk -- HD 100546 --- a low gas/dust ratio is inconsistent with observed CO line fluxes, and instead suggest that the disk gas is depleted in carbon.  A natural way to preferentially deplete carbon in the disk gas is to convert it into a less volatile form.  This idea is also supported by the first attempt to directly measure the CO/H$_2$ ratio in a disk; using observations of HD and CO in the TW Hya disk, \citet{Favre13} derive a CO/H$_2$ ratio a few to 100 times lower than the canonical value, suggesting the conversion of much of the carbon in CO into more refractory compounds, such as organics.

Protoplanetary disk observations have also been used to search for evidence for isotope-selective photodissocation of CO, predicted to be the cause of the solar system's mass-independent oxygen anomalies (see Section \ref{sec:fractionation}).  Using high-resolution spectrosopic observations of protoplanetary disks, \citet{Smith09} directly measured the abundances of C$^{16}$O, C$^{17}$O and C$^{18}$O. They found evidence for mass-independent fractionation, consistent with a scenario in which the fractionation is caused by isotope selective photodissocation in the disk surface due to CO self-shielding \citep{Lyons05,Visser09}.

\section{\textbf{Disk volatiles and exoplanets}}

In Section \ref{SolarNebula}, we discussed how the presence of a water snow line shaped the composition of the different planets and other bodies in the solar system, depending on their formation location and, perhaps, dynamical history. We also demonstrated that molecular species other than water must have had similar strong radial gas-to-solid gradients, as preserved in elemental and isotopic abundances. Now that we have an increasingly detailed, if not yet fully comprehensive, picture of protoplanetary disk chemistry and thermodynamics, supported by astronomical data, it is natural to ask how volatiles impact the architecture and composition of exoplanets. Planetary properties that may be influenced by the natal volatile composition include system characteristics, such as the distribution of masses and orbital radii, their bulk chemistry (the fractions of rock, ice and gas) and their atmospheric and surface chemistry.  

While the combination of inner, dry terrestrial planets and outer icy giants and planetesimals in the solar system strongly suggest that the water snow line is irreversibly imprinted on the large-scale architecture of a planetary system, the discovery of hot-Jupiter planets on short-period orbits showed that there is, in fact, no such {\it universal} relationship \citep{Mayor95, Marcy96}. The solution does not appear to be to allow giant planets to form close to the parent star (well within 1 AU) -- the consensus is that this is not possible because sufficient amounts of mass cannot be maintained within such small radii. Rather, the planets migrate radially through some combination of gas interactions during the protoplanetary disk phase \citep{Lin96}, dynamical interactions with a planetesimal cloud \citep{Hahn99}, and  three-body interactions with a stellar companion \citep{Fabrycky07}. While a discussion of the dynamical evolution of planetary systems is beyond the scope of this chapter, the important point, when relating exoplanets to the volatile system in protoplanetary disks, is that observed planetary orbits cannot be assumed to reflect their birth radii and feeding zones. The chemical composition of their planets may have been set by disk conditions that were never present at their ultimate mean orbital radius. In the solar system, it was difficult to deliver the Earth's water, while exo-planetary systems may have common water worlds that formed beyond the snow line, but heated up either due to orbital or snow line migration \citep{Tarter07}.

Planetary migration may be seen as a complication when interpreting observations, but perhaps it is also an opportunity. We review two different aspects relating disks to exoplanets. One is the question of whether exoplanet architecture (bulk mass and orbital structure) reflects the properties of the volatile distribution in their parent disks, even though some planetary migration takes place. Another is whether the chemical structure of exoplanet atmospheres constrain their birth radii, and therefore provides an independent measure of the migration process by allowing a comparison between the initial and final orbital radii.

\vspace{0.15cm}
\subsection{The importance of solid volatiles in planet formation}
\vspace{0.15cm}

Do condensible volatiles play a central role in the basic ability of protoplanetary disks to form planets of all types, or are they merely passengers as the rocky material and H$_2$/He gas drive planet formation? The answer to this question generally seems to be that condensible volatiles do play a central role. As discussed in section \ref{IceMass}, the total mass of condensible volatiles available for planetesimal formation is at least twice that of rock, and maybe as much as 3--4 times as much, at radii beyond the main condensation fronts. The large increase in solid mass offered by condensing volatiles affects the growth of planets on all scales -- in grain-grain interactions, in the formation of planetesimals and in the formation of gas giants. 

At the microscopic level, icy grains may stick more efficiently at higher collision velocities, than bare rocky grains. Simulations suggest that the break-up velocity for icy grains may be 10 times higher than that of silicate grains \citep{Wada09}. Other mechanisms that aid in the growth of icy grains include the potential to melt and rapidly re-freeze upon high velocity collisions, leading to sticking even in energetic collisions \citep{Wettlaufer10}.

An increase in solid mass may also catalyze the formation of planetesimals -- a critical step without which no planet formation could take place in any but the most massive disks. At sufficiently high solid surface densities, the so-called streaming instability can lead to the rapid formation of planetesimals by gravitational collapse. This is due to a combination of pressure gradient flows and drag forces exerted by solid particles back onto the gas \citep[e.g.,][]{Youdin02, Johansen07}. 

In the context of giant planet formation by core accretion, \cite{Pollack96} demonstrated that a factor of 2 in solid surface density can lead to a reduction in giant planet formation times by a factor 30. The time scale is determined by the time it takes for a planet to reach the crossover mass -- the mass at which runaway gas accretion occurs. Since the lifetime of the gas in protoplanetary disks is comparable to the time scale for effective core accretion, factors of a few in solid mass can make a large difference in whether a system forms giant planets or not. Planet formation models that include more complexity, such as orbital migration \citep{Ida04a}, and evolving disks across a wide parameter space \citep{Mordasini09}, confirm this basic conclusion and emphasize the need for solid-state volatiles around solar-metallicity stars to form gas giants. 

The dependence of planet formation on the availability of solid mass is famously reflected in the positive relation between stellar metallicity and the frequency of gas giants with masses $\gtrsim M_{\rm Jup}$ \citep{Santos04,Fischer05}, coupled with a lack of such a relation with Neptune-mass planets \citep{Sousa08}. The interpretation of these observations is that low metallicity disks cannot form giant planets within the disk life time because the solid cores never grow sufficiently massive to accrete large gas envelopes, while the Neptune-mass planets can form over a wider range of stellar metallicity \citep{Ida04b}.

\vspace{0.15cm}
\subsection{Condensation fronts: A generalized concept of snow lines}
\vspace{0.15cm}
\label{sec:condensation}

When searching for links between protoplanetary disk chemistry and exoplanets, it becomes necessary to generalize the concept of the snow line. Any molecular, or even atomic, species with condensation temperatures between 10 and 2000\,K will have a curve demarcating the border beyond which the partial vapor pressure drops dramatically (the species freezes out). Further, the border is rarely a straight line, but a surface that roughly follows an isotherm, with some departures due to the pressure dependency on the condensation temperature. In a disk, a condensation front is typically strongly curved and becomes nearly horizontal for disk radii dominated by external irradiation heating \citep{Meijerink09}, or as a pressure gradient effect in regions that are mostly isothermal \citep{Ros13}. Use of the term {\it condensation front} to describe this surface for a given molecular or atomic species allows the snow line nomenclature to be retained as a specific reference to water freeze-out at the disk mid plane, a distinction we will utilize in this text.

\begin{figure}[ht!]
\includegraphics[width=8cm]{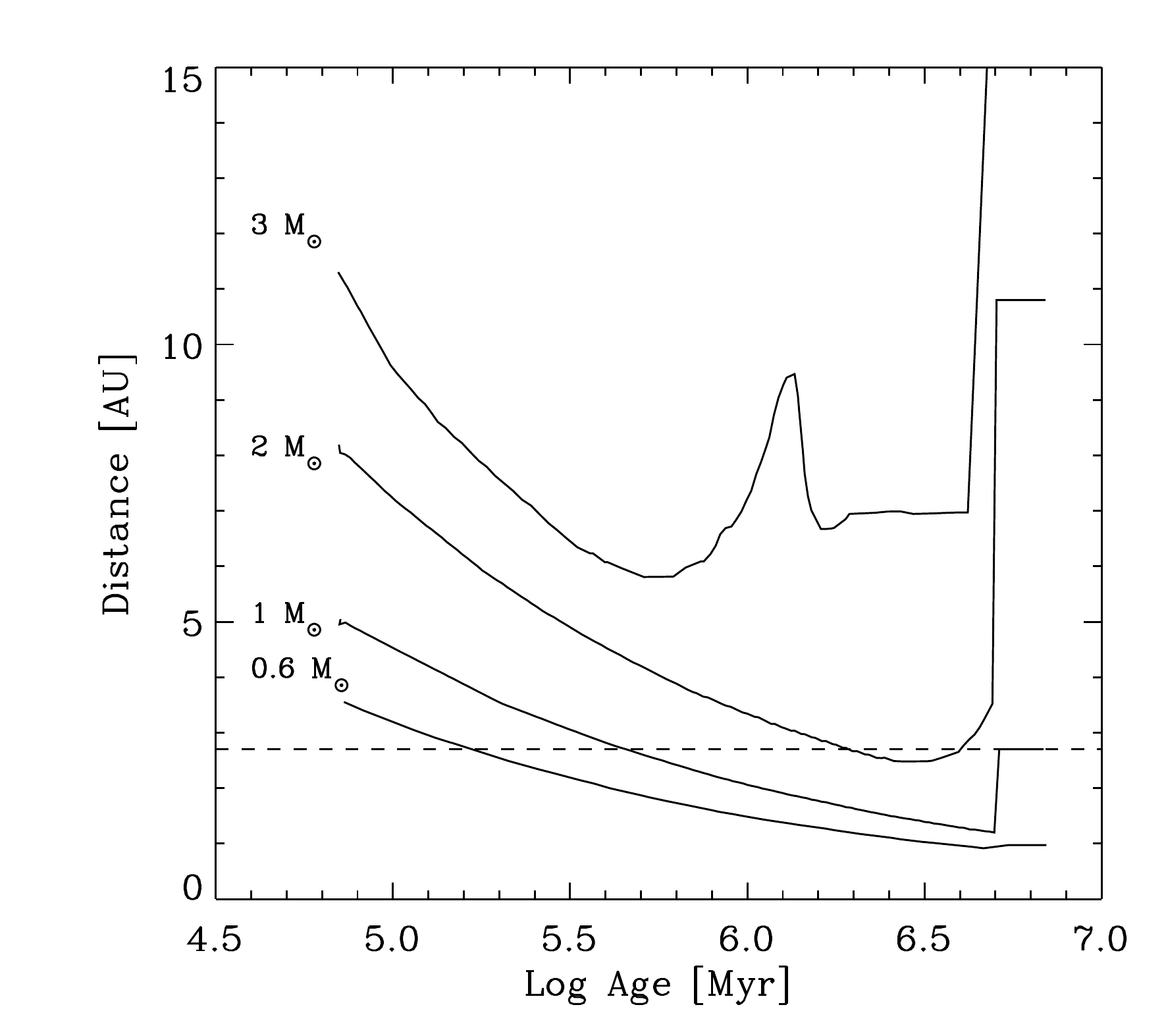}
\caption{The evolution of the midplane snow line as a function of stellar mass. The data are taken from \cite{Kennedy08} and modified with an optically thin end stage due to disk dissipation.  Aside from the final optically thin stage, the shapes of the curves are determined by the stellar luminosity and accretion history of the disk. Note that the 3 M$_\odot$ star undergoes large changes in luminosity as it approaches the main sequence, while the lower-mass stars never reach the main sequence during the disk lifetime. The dashed line represents the predicted snowline for an optically thin disk around a sun-like star, originally derived by \citet{Hayashi81}.}
\label{fig:snowline}
\end{figure}

Condensation fronts evolve strongly with time as the disk temperature changes in response to evolving stellar insolation and internal disk heating \citep[e.g.][]{Lissauer87}, and with stellar luminosity \citep{Kennedy08}. At the earliest stages, high accretion rates can push the snow line out to 5 AU or more around a young solar mass star. At intermediate stages (1-2 Myr) during which giant planets may accumulate most of their mass, disk accretion rates drop toward $10^{-9}\,M_{\odot}\,\mathrm{yr}^{-1}$, decreasing viscous heating to the point where the snow line moves within 1 AU. At the latest stages of evolution, the disk clears, and the snow line radius is again pushed outwards as the disk becomes optically thin in the radial direction exposing the disk midplane to direct stellar irradiation \citep{Garaud07,Zhang13}. We show the predicted evolution of midplane snow line distances for a range of stellar masses in Figure \ref{fig:snowline} (based on the models of \citealp{Kennedy08}). Importantly, it is seen that, while the end-stage snow line is located well outside the habitable zone of a solar mass star \citep[where liquid water may exist on a planetary surface;][]{Kasting93}, it was actually inside the habitable zone during the critical period in the planet formation process that sees the formation of giant planets and the planetesimal system. Consequently, there is a possibility that habitable water worlds are more common among exo-planetary systems than the solar system would initially suggest. 

We can also consider the dependence of stellar mass on this basic form of the snow line evolutionary curve. While stars with $M_*\gtrsim 1\,M_{\odot}$ have disks that during the optically thick phase experience a snow line radius significantly closer to the star than the optically thin end stage configuration, disks around lower-mass stars follow a different path. Around these stars, the optically thin end-stage snow line is the closest it gets. This is due to the relative dominance of accretion heating versus direct stellar irradiation around the low luminosity low-mass stars, suggesting that planetary migration is more important for forming water worlds in low-mass systems.  

\vspace{0.15cm}
\subsection{Transport of volatiles}
\vspace{0.15cm}

Protoplanetary disks are evolving systems that transport a variety of quantities, angular momentum, mass and energy \citep[see][for a general review]{Armitage11}. The transport and mixing of solids in protoplanetary disks is a large, general area of study, but one that has important consequences for disk volatiles. Like refractory material, condensed-phase volatiles can move, in bulk, relative to each other and to the primary H/He gas reservoir. However, as opposed to the refractories, a property of volatiles that make them a special case is their ability to co-exist in solid and gaseous forms and to go through rapid phase changes across relatively sharp condensation fronts. Two different transport mechanisms may act in conjunction to concentrate volatiles, and they are therefore often discussed together: The movement of gasous volatiles by diffusion across the steep partial pressure gradients of a condensation front and the migration of boulder-sized icy bodies against a shallow bulk (H/He) gas pressure gradient. 

Radial pressure gradients in protoplanetary disks are well-known catalysts of solid transport through advection. In the absence of other effects, they are sufficiently efficient to move large fractions of the entire solid mass in the disk. Such pressure gradients support the gas by a radial force component, allowing it to orbit at slightly sub- or supra-Keplerian velocities, depending on the direction of the gradient. Gas that is not in a Keplerian orbit, in turn produces a dissipative force on solid particles, causing them to lose or gain angular momentum and migrate radially \citep{Weidenschilling77,Stepinski97}. The rate of advection is a sensitive function of particle size, with the highest rates seen for sizes of 1-100\,cm. For viscously evolving disks, the pressure gradient is monotonically decreasing with radius, leading to rapid inwards advection. This mechanism has the power to dramatically change the surface density of, in particular, volatiles, as icy bodies are transported from the outer disk toward, and across the snow line \citep{Ciesla06}. 

As the ices cross over their respective condensation fronts and subsequently sublimate, the partial pressure of each volatile species sharply increases. Such steep partial pressure gradients are not stable, and the volatile gas will diffuse out across the condensation front again where it re-freezes (mostly due to turbulent mixing, although other types of mixing may also operate). In the absence of inwards advection, this can lead to the complete depletion of inner disk volatiles. The process of diffusion against a partial pressure gradient is known as the {\it cold finger effect} \citep{Stevenson88}.  Together, these effects -- inwards solid advection and outwards gas diffusion -- may, for certain disk parameters, increase the solid volatile surface density just beyond the condensation fronts by an order of magnitude \citep{Cuzzi04,Ciesla06}.

An important consequence of the transport of volatile species toward condensation fronts located at different disk radii may be to impose a strong radial dependence on the elemental C/O ratio, as suggested by models of meteorite formation \citep{Hutson00, Pasek05}. In direct analog to other astrochemical environments where the elemental C/O ratio dictates a water- or carbon-dominated chemistry, disks are expected to be similarly sensitive. It may therefore be possible to measure the disk C/O ratios at various radii by measuring the relative abundances of water and carbon-bearing volatile species. A range of HCN/H$_2$O ratios measures using Spitzer spectra \citep{Salyk11, Carr11}, and an observed correlation of HCN/H$_2$O line strengths with disk mass is at least consistent with the possibility that planetesimal growth could sequester O at or beyond the snow line and increase the inner disk C/O ratio \citep{Najita13}. 

Vertical transport of volatiles, while involving less bulk mass than radial transport, may act to significantly alter the observables of disk volatiles. In the range of radii in the disk where the condensation fronts are nearly horizontal (and very close in height) due to direct stellar irradiation of the disk, the cold finger effect may act in the vertical direction. That is, volatile vapor turbulently mixes across the condensation front, below which it freezes out. Some of the resulting solids will then settle toward the midplane, leading to a net loss of volatiles from the disk surface. This was used as a proposed, but qualitative, explanation for the observed water vapor distribution in the AS 205N disk, in which Spitzer spectroscopy suggested a snow line near 1 AU, well within the $\sim 150\,$K region normally demarcating the snow line \citep{Meijerink09}. The effect was subsequently supported by the numerical models of \cite{Ros13}. 

Beyond the condensation front, grain settling may also deplete the surface of all solid volatiles. This was suggested to explain the extremely low abundances of water vapor observed with HIFI onboard the Herschel Space Observatory in the outer TW Hya disk ($\gtrsim 50\,$AU) \citep{Hogerheijde11}. While it may seem counter-intuitive to constrain the abundance of water ice using water vapor, this works in the disk surface because ultra-violet and X-ray radiation will desorb water ice into the gas-phase at a constant rate at a given incident intensity \citep{Ceccarelli05, Dominik05}. In the case of TW Hya, the water vapor abundance was found to be 1-2 orders of magnitude lower than expected from photo-desorption, leading to a suggestion that the water ice was depleted by a similar factor through a process of preferential settling. Similarly low amounts of water vapor are consistent with upper limits measured by \citet{Bergin10} around DM Tau.

\vspace{0.15cm}
\subsection{Observations of condensation fronts}
\vspace{0.15cm}

Based on the development of high fidelity mid-infrared spectroscopy targeting warm volatiles, including water, as well as some of the first ALMA line imaging results, it has now become feasible to measure, and even image, the location of condensation fronts. The observations address specific technical challenges in measuring a condensation front. 

One challenge is related to the small angular size of the gas-rich region (10s of milli-arcseconds for water and $\sim$0.5 arcseconds for CO in nearby star forming regions); to measure the snow line with a direct line image, the spatial resolution of the observation must be significantly better than these sizes. To characterize the shape and sharpness of the condensation front, the resolution must be even higher. As an alternative to direct imaging, the many transitions of water covering virtually all excitation temperatures provides an opportunity to spectrally map its radial distribution in a disk, using temperature as a proxy for radius. The first qualitative attempt was made by \cite{Meijerink09} for the disk around AS 205N, and the method was refined and quantified for the transitional disk TW Hya by \cite{Zhang13}.  

In TW Hya, the inner few AU are depleted of small dust grains \citep{Calvet02}, and the snow line is located further from the star than would be expected for a classical optically thick disk. Instead, the decreased radial optical depth of grains allows radiation to penetrate to larger radii in the disk, and moves the snowline to just beyond the transition radius where the disk becomes optically thick. This observation is evidence that the snow line location moves outwards near the end of the gas-rich, optically thick phase of protoplanetary disk evolution. 

One challenge to the temperature mapping method is that the disk midplane is shielded, either by highly optically thick dust or by interfering line emission from a superheated surface layer. For condensation fronts at small radii and relatively high temperatures, such as for water, the relevant mid- to far-infrared lines are therefore formed in the disk surface above one scale height \citep{Woitke09}, as opposed to in the midplane. In practice, this may skew measured condensation fronts to larger radii.

\begin{figure}[ht!]
\includegraphics[width=8cm]{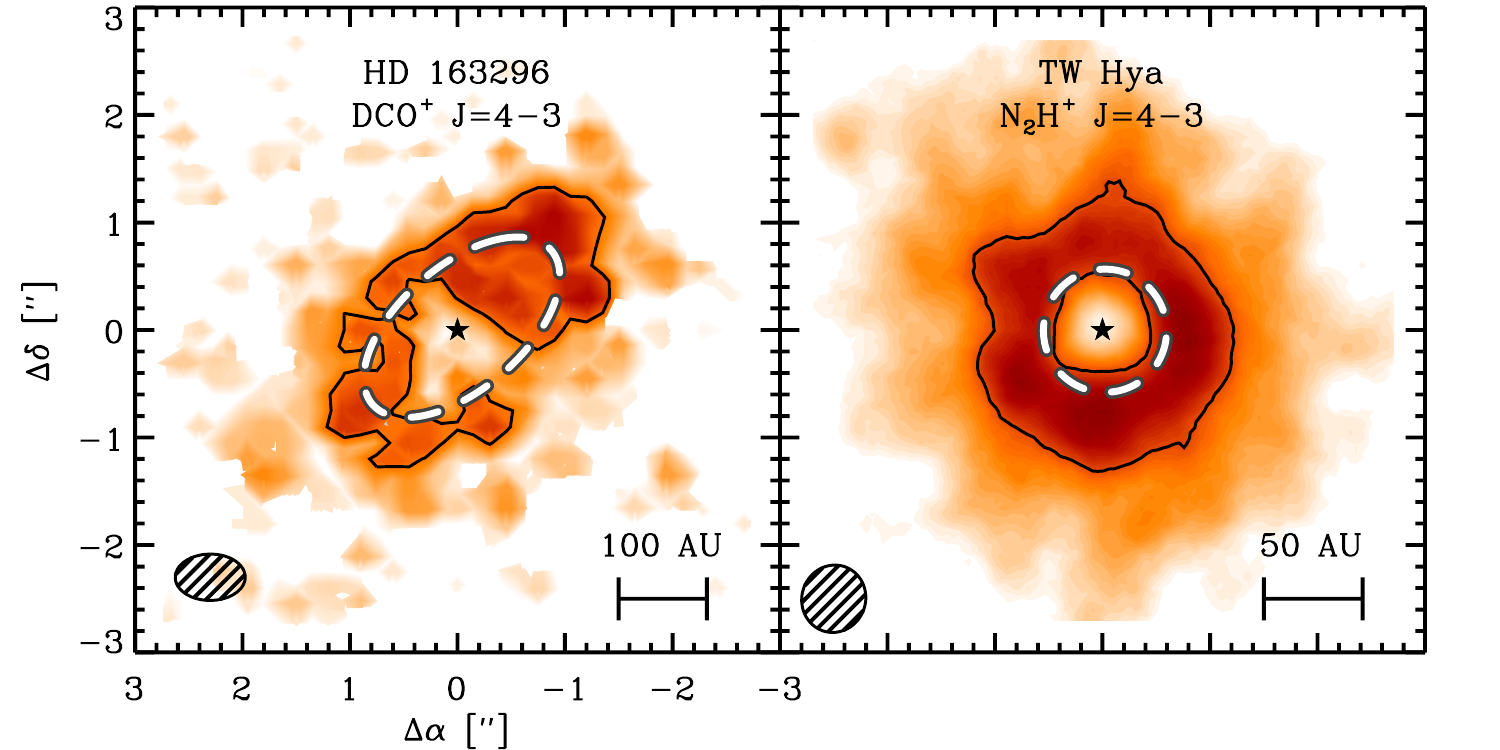}
\caption{The CO snowline as observed with ALMA. The CO snowline is observed using chemical tracers of the absence of CO from the gas-phase (\citealp[left, DCO$^+$][]{Mathews13} and \citealp[right, N$_2$H$^+$][]{Qi13b}). The dashed curves delineate the 17\,K isotherms, where CO freezes out.}
\label{fig:obs_co_snowline}
\end{figure}

While water has the most famous condensation front, others, characterized by lower temperatures, are large enough to image directly with facilities like ALMA.  Evidence for CO freeze-out has been previously suggested due to a measured overall depletion of CO in disks relative to dark cloud values \citep[e.g.][]{Qi04}. However, attempts to image the CO condensation front directly in disks using rotational CO lines have proven to be more challenging than initially expected \citep{Qi08}. The CO condensation front tends to get washed out in line images due to the high opacity of low-$J$ CO lines and filling-in by warm surface gas. Effectively, the disk atmosphere hides the depletion in the disk midplane. The solution to this problem has been to use {\it chemical} tracers. That is, molecules that only exist in abundance when the CO is absent from the gas phase and/or abundant in the solid phase, thus avoiding interference by emission from CO-rich regions. Four molecules have been suggested: N$_2$H$^+$, which is efficiently destroyed by gas-phase CO \citep{Qi13a}, H$_2$CO and CH$_3$OH, which form from CO ice hydrogenation, and DCO$^+$, which is enhanced as CO depletion allows the formation of one of its parent molecules,  H$_2$D$^+$ \citep{Mathews13}.   Examples of CO condensation fronts measured using DCO$^+$ (HD 163296) and N$_2$H$^+$ (TW Hya) are shown in Figure \ref{fig:obs_co_snowline}.  

N$_2$H$^+$, H$_2$CO and CH$_3$OH should all be present exclusively outside of the CO snow line, with an inner edge that traces its position.  The case of DCO$^+$ is complicated because CO is also a parent molecule, and its abundance is a result of a competition between CO and H$_2$D$^+$, leading to efficient formation in a narrow temperature range just around the CO condensation front. Recent images of DCO$^+$ in the disk around HD 163296 show a ring-like structure, with a radius consistent with that expected for CO condensation, suggestive that at least for this disk, DCO$^+$ is a good CO snow line tracer \citep[][Figure \ref{fig:obs_co_snowline}]{Mathews13}. However, TW Hya does not show the same ring-like structure in  DCO$^+$ \citep{Qi08} while a more recent ALMA image of N$_2$H$^+$ in this disk confirms a clear ring-like structure, with a location consistent with the expected radius of the CO snow line \citep[][Figure \ref{fig:obs_co_snowline}]{Qi13b}. 

\vspace{0.15cm}
\subsection{The influence of disk volatiles on planetary composition}
\vspace{0.15cm}

It is a basic expectation that the bulk elemental composition of a planet reflects that of protoplanetary disk material that formed it. Further, as planets never fully mix and equilibrate, interiors and atmospheres may also reflect differences in the composition of material accreted during different stages of formation. This dependence is supported by the wide diversity of planets (and likely, moons) in exoplanetary systems; clearly, not all planets are the same, and to what degree is this due to disk volatiles and the processes that sculpt their distribution in a protoplanetary disk? We are just beginning to be able to study the compositions of exoplanets as well, and diversity appears to be the rule, rather than the exception. Thanks to the success of the Kepler mission \citep{Borucki10}, other exoplanet transit studies \citep[e.g.,][]{Barge08} and radial velocity surveys \citep{Mayor03,Cumming08} combined with transit observations, we can now determine the average density of some exoplanets, and study their bulk chemistry \citep[e.g.,][]{Sasselov03,Burrows07}.  

Further, the first observations of the atmospheric compositions of exoplanets are being made possible with studies of primary and secondary planetary transits \citep{Knutson07}, as well as direct imaging \citep{Konopacky13}.  However, the relatively small number of planets that have been studied have atmospheric signatures that are not trivially reproduced by existing atmospheric models \citep[e.g.,][]{Barman11, Skemer12}.

One important parameter, among others, in modeling exoplanet atmospheres is the elemental C/O ratio. The possible detection of a high C/O ratio in the exoplanet WASP-12b prompted a study suggesting a simple way in which the disk architecture might affect the composition of planetary atmospheres \citep{Madhusudhan11}. \citet{Oberg11b} suggested that, if a giant planet atmosphere does not mix, or mixes incompletely with volatiles sequestered in its core, it is possible to create an atmosphere with a high C/O ratio near unity. This is possible in regions of the disk where oxygen preferentially condenses relative to carbon (for instance outside the water condensation front, but inside the CH$_4$, CO$_2$ and CO fronts). If planets form via core accretion from solids followed by the accretion of a gaseous envelope, their atmospheres will inherit the high C/O ratio, driving an atmospheric carbon-dominated chemistry. This predicts different planetary atmospheric compositions as a function of initial formation radii, but independently of the final planetary orbits.

\section{\textbf{The physical structure and kinematics of protoplanetary disks as traced by volatiles}}
\vspace{0.15cm}

Our knowledge of the physical structure and kinematics of protoplanetary disks is fundamentally dependent on how well we understand the behavior of their volatiles. By their nature, volatiles can be found in the gas-phase throughout the disk. Indeed, the total mass of protoplanetary disks is entirely dominated by a volatile molecule -- H$_2$. As we have seen in the previous sections, much of the most abundant heavy elements are also often found in the form of gas-phase volatiles. This is fortunate because, in their absence, our current knowledge of the physical structure and kinematics of protoplanetary disks would be far less advanced; we rely on molecular line tracers to measure velocities, temperatures and mass of the gas. The problem with using volatiles as physical tracers arises because the disks are far from well-mixed, chemically simple systems. Thus, kinematic, thermal and abundance measurements derived from molecular lines depend on a model of the volatile system; are most of the tracer molecules frozen out? Are we tracing a specific layer where chemical formation happens to be more effective? Have the total gas+solid volatile reservoir been depleted due to transport?

In this section, we discuss the links between the physi-chemical behavior of disk volatiles and their use as physical tracers, and in particular point out areas where our improved understanding of volatiles improve the physical measurements. A more general discussion of measurements of physical properties of protoplanetary disks can be found elsewhere in this book. 

\vspace{0.15cm}
\subsection{The impact of volatiles on disk mass measurements}
\vspace{0.15cm}

The bulk gas masses and the gas-to-dust ratios of protoplanetary disks are some of the most fundamental observational parameters. They are often used to place the disks into a global evolutionary scheme, or to compare their properties to those inferred for the earliest stages of solar system formation. 

The bulk mass of disks has most commonly been measured by observing the dust mass and converting to a gas mass.  More recently, attempts have been made to measure the gas mass directly, via various gas-phase tracers.  However, no matter the method of bulk mass measurement, an understanding of the volatile content of the disk is required, as the volatiles represent most of the disk mass.  Therefore, many of the intrinsic properties of volatiles discussed elsewhere in this chapter also create difficulties for attempts to measure bulk disk masses.  Nevertheless, there has been significant progress on this front, which we describe here.

The use of the dust continuum emission is the oldest method to estimate the total disk mass, and is still the most frequently used \citep{Beckwith90,Andrews05,Mann10}. In converting from dust mass to total disk mass, the gas-to-dust ratio and dust opacities are typically assumed. However, both of these assumed values are traditionally based on dust models that often do not include any of the CNOH volatiles. Are these assumptions actually reasonable, given the knowledge that the disk dust mass is dominated by ices? For disk material in the limiting (but well-mixed) case that everything, except for hydrogen and helium, is in the solid state, the solar abundances of \citet{Grevesse10} imply a gas-to-dust mass ratio of 72. The consequence is that disk masses are probably slightly lower than if the gas-to-dust ratio is assumed to have the canonical value of 100, but only by a relatively small fraction. In terms of the dust opacity, the addition of ices has a complex behavior due to several competing effects. Ice itself generally has lower mass opacity than refractory dust, so the opacity of bulk disk material will decrease significantly with the addition of ice. Secondary effects, such as a potentially increased sticking efficiency may change the opacity of icy grains by catalyzing grain growth. All together, the net effect seems to be to slightly decrease the mass opacity of ice by 20--30\% \citep{Ossenkopf94,Semenov03}. Thus, volatiles actually have a relatively minor effect on determination of bulk dust masses in disks. The one potential caveat is if ices convert from simple species to complex CNOH organics, as suggested by solar system meteoritics. The organics have much higher opacities than, say water, CO$_2$ or NH$_3$ ice, and may dominate the total dust opacity at high abundances \citep{Pollack94}. 

One approach to estimate disk masses with gas-phase tracers has been to observe multiple rotational transitions of multiple isotopologues of CO with millimeter interferometers.  The more optically thin isotopologues are sensitive to the disk mass, while the optically thick transitions (especially from higher-excitation states) are sensitive to the disk temperature structure. Using this method, \citet{Panic08} simultaneously measure the gas and dust masses in the disk around HD 169142 and find a gas/dust ratio in the range of 25--100.  

A new method for estimating disk mass was explored by \citet{Bergin13}, who used observations of HD $J=1-0$ obtained with Herschel-PACS to measure the disk mass around TW Hya. HD represents an attractive mass tracer for two main reasons: It directly probes the bulk molecular component (H$_2$) with a conversion factor from HD to H$_2$ expected to be near the interstellar value for nearly all regions of the disk, and, HD, unlike CO, is not expected to freeze out in disk conditions. However, as for all gas tracers, the result is still model-dependent through the assumed gas temperature. \citet{Bergin13} used observations of CO millimeter transitions to independently constrain the gas temperature distribution in the disk. They found a disk mass consistent with a gas/dust ratio near 100, suggesting that even this evolved disk has a canonical gas/dust ratio and sufficient mass to form giant planets. Measurements of the gas mass in TW Hya were also attempted by \citet{Thi10} and \citet{Gorti11}, both of whom used a large suite of emission lines, including millimeter-wave CO lines, to constrain the disk properties. However, the model dependencies in this approach were exposed by the discrepant results produced by the two groups, which allowed for TW Hya to have either a very sub-canonical or nearly canonical gas/dust ratio. Use of the method presented by \citet{Thi10} also gives gas-to-dust ratios significantly below 100 for other disks \citep[e.g.,][]{Tilling12}, demonstrating a strong need to reach a consensus using different methods for disk mass measurements.

\vspace{0.15cm}
\subsection{The kinematics of disk volatiles}
\vspace{0.15cm}

The bulk mass of protoplanetary disks resides close to the midplane, and can be described by a Keplerian flow; known departures from this, for instance due to radial pressure support \citep{Weidenschilling77} are very slight ($\lesssim 0.01 V_{\rm Kepler}$), and still beyond our observational capabilities. The surface layers of the disk, however, may be subject to much greater dynamical effects in the form of enhanced ionization-dependent turbulence and winds. Some chemical models now begin to consider the efficiency of processing volatiles as they are mixed into a turbulent surface layer and exposed to high temperatures and direct irradiation before being returned to the midplane \citep{Semenov11,Furuya13}. As all chemical time scales in the midplane are much longer than the photo-chemical time scale on the disk surface, such mixing can lead to significant changes to the bulk chemistry, if efficient.

Volatile emission lineshapes also hint at the tidal influence of planets.  Models of forming gas giant planets embedded in circumstellar disks indicate that their interactions cause the inner edge of the cleared region exterior to their orbit to grow eccentric \citep{Papaloizou01, Kley06, Lubow91}. When the companion mass grows to more than 0.3\% of the mass of the central star, simulations show that azimuthally averaged eccentricities can grow as large as $e=0.25$ at the inner edge of the disk yet fall off as fast as $r^{-2}$ \citep{Kley06}. Gas arising from material in a Keplerian orbit with a non-zero eccentricity reveals a distinct shape \citep{Liskowsky12} due to the uneven illumination of the inner edge of the disk and a slight Doppler shift relative to a circular orbit due to the non-circular velocities. \citet{Regaly10} describe the ro-vibrational CO line profiles that could arise from a disk harboring a  supra-jovian mass companion.  However, the CO emission originates from a large radial extent, diluting the signal from the asymmetric inner rim \citep{Liskowsky12}.  A better tracer of the inner rim is ro-vibrational OH, which may be excited preferentially in a ring at the inner edge of the outer disk.  Such asymmetric ro-vibrational OH emission has been observed toward V380 Ori \citep{Fedele11} and HD~100546 \citep{Liskowsky12}.  It should be noted that other scenarios could produce asymmetries; for example, fluctuations of turbulence \citep[e.g.][]{Fromang06} could give rise to non-axisymmetric density structures that result in asymmetric emission from OH, which originates in a region near the critical density for thermalizing its vibrational levels.  Further temporal studies of this effect are in progress to distinguish between the possible scenarios.

\section{\textbf{Future directions}}
\vspace{0.15cm}

Our understanding of protoplanetary disk evolution and planet formation is in a transitional stage. Where low resolution observations once led to models of disks as monolithic objects, characterized by a few scalar parameters (mass, size, dust grain size, etc.), disks are now seen as well-resolved composite entities with different, highly physically and chemically distinct regions. We are no longer averaging properties over entire disks, but are taking the first steps toward deconstructing them by zooming in on specific sub-regions to derive their causal interrelations. In this chapter, we have shown how volatiles play a key role in sculpting a changing environment as we move from the inner terrestrial region of protoplanetary disks, across the giant planet region and toward the Kuiper belt and cometary regions. We expect that during the next 5-10 years, we will see dramatic progress in our understanding of disk chemistry, the transport of bulk volatiles and their links to exo-planetary systems:

\begin{itemize}
\item{{\bf Elemental budgets in disks:} One of the most direct, and observationally feasible, comparisons of protoplanetary disks to the conditions in the solar nebula, is the distribution of elements in volatile and refractory molecules. We expect that significant progress in detecting all the bulk elemental volatile carriers throughout protoplanetary disks will be made in the next years. However, the success of this task will depend not only on new data, but also on improving the accuracy with which radiative transfer calculations can retrieve the abundances. An important question is whether our inability to observe N$_2$ will result in the nitrogen budget remaining elusive for the foreseeable future?}
\item{{\bf Inheritance or reset?} A central question is that of inheritance. To what degree is the volatile composition transferred into disks from the dense interstellar medium through the protostellar phase, and to what degree are the volatiles formed in situ within chemically active regions in the disk, resetting previous chemical signatures and losing memory of the interstellar phase. Likely, the answer will be that both types of evolution play a role with the latter being more important for the inner disk. Improvements in the accuracy with which chemical abundances can be retrieved as a function of location in the disk, with spatial fidelities as high as a few AU are required. This will be provided by a synergistic combination of submillimeter interferometric line imaging with ALMA with mid- to far-infrared spectroscopy with JWST and a range of ground-based optical/IR telescopes. Importantly, no single observatory will be sufficient in isolation because of the need to cover a wide range in gas temperatures and corresponding wavelength ranges.}
\item{{\bf Transport of volatiles:} Models have long predicted that the local concentration of volatiles may vary by orders of magnitude in different disk regions due to a combination of gas diffusion and solid particle advection. Collectively, these transport processes may alter the observational signature of the disk surface, dramatically change the local mode of chemistry by changing the C/O ratio and increase the local solid surface density to aid in the formation of planetesimals and planets. The near future is likely to bring observational searches for bulk transport of volatiles that test dynamo-chemical models of protoplanetary disks.} 
\item{{\bf Planet composition as a function of volatile content:} The composition of individual planets must necessarily reflect that of the disk material used in their formation. Thus, there is an expectation that the chemical demographics of disks will match the compositional demographics of exoplanetary systems. The available data are still far too insufficient to make such connections, but the rapid advances in both disk and planet observations are likely to soon make such comparisons common.} 
\end{itemize}

\textbf{ Acknowledgments.} This work was supported by a NASA Origins of the Solar System Grant No. OSS 11-OSS11-0120, a NASA Planetary Geology and Geophysics Program under grant NAG 5-10201, and by the National Science Foundation under grant AST-99-83530. The authors are grateful to Ewine van Dishoeck and Thomas Henning for comments that greatly improved this chapter.  

\bibliographystyle{ppvi_lim1}
\bibliography{volatiles_ppvi}

\end{document}